\documentclass[aoas, preprint]{imsart}

\RequirePackage[OT1]{fontenc}
\RequirePackage{amsthm,amsmath}
\RequirePackage[]{natbib}
\RequirePackage[colorlinks,citecolor=blue,urlcolor=blue]{hyperref}
\RequirePackage{graphicx}
\RequirePackage[]{algorithm2e}

\startlocaldefs
\numberwithin{equation}{section}
\theoremstyle{plain}

\endlocaldefs

\usepackage{amssymb}
\usepackage{graphicx,color}
\usepackage{amsfonts}

\usepackage{booktabs,bm,multirow, pifont,multirow, placeins}

\newcommand{\ttt}{{\boldsymbol\theta}}

\newcommand{\aaa}{{\boldsymbol\alpha}}

\newcommand{\aalpha}{{\boldsymbol\alpha}}

\def\AA{\mbox{$\mathbf A$}}

\newcommand{\dd}{{\mathbf d}}
\newcommand{\DD}{\mbox{$\mathbf D$}}

\def\gg{\mbox{$\mathbf g$}}

\newcommand{\pp}{\mbox{$\mathbf p$}}

\newcommand{\qqq}{{\mathbf q}}

\newcommand{\RR}{{\mathbf R}}

\def\sb{\mbox{$\mathbf s$}}

\newcommand{\zz}{\mbox{$\mathbf z$}}

\begin{document}

\begin{frontmatter}
\title{Partial-Mastery Cognitive Diagnosis Models}
\runtitle{Partial-Mastery Cognitive Diagnosis Models}


\begin{aug}
\author{\fnms{Zhuoran} \snm{Shang}\ead[label=e1]{zhuoranshang@gmail.com}},
\author{\fnms{Elena A.} \snm{Erosheva}\ead[label=e2]{erosheva@uw.edu}}
\and
\author{\fnms{Gongjun} \snm{Xu}\ead[label=e3]{gongjun@umich.edu}}


\runauthor{Shang et al.}

 \address{ 
 School of Statistics,
  University of Minnesota
\printead[presep={,\ }]{e1}}
  
 \address{Department of Statistics, School of Social Work,    the Center for Statistics and the Social Sciences, 
  University of Washington \printead[presep={,\ }]{e2}}

 \address{  Department of Statistics,
 University of Michigan
\printead[presep={,\ }]{e3}}
 
\end{aug}

\begin{abstract}
Cognitive diagnosis models (CDMs) are a family of discrete latent attribute models that serve as statistical basis in educational and psychological cognitive diagnosis assessments. CDMs aim to achieve fine-grained inference on individuals' latent attributes, based on their observed responses to a set of designed diagnostic items. In the literature, CDMs usually assume that items require mastery of specific latent attributes and that each attribute is either fully mastered or not mastered by a given subject. 
We propose a new class of models, partial mastery CDMs (PM-CDMs), that generalizes CDMs by allowing for partial mastery levels for each attribute of interest. We demonstrate that PM-CDMs can be represented as restricted latent class models. Relying on the latent class representation, we propose a Bayesian approach for estimation. We present simulation studies to demonstrate parameter recovery, to investigate the impact of model misspecification with respect to partial mastery, and to develop diagnostic tools that could be used by practitioners to decide between CDMs and PM-CDMs. We use two examples of real test data -- the fraction subtraction and the English tests -- to demonstrate that employing PM-CDMs not only improves model fit, compared to CDMs, but also can make substantial difference in conclusions about attribute mastery. We conclude that PM-CDMs can lead to more effective remediation programs by providing detailed individual-level information about skills learned and skills that need to study.  
\end{abstract}

%
\begin{keyword}
\kwd{Cognitive diagnosis}
\kwd{Restricted latent class models}
 \kwd{Mixed Membership models}\\
 {This work has been published in Ann. Appl. Stat. 15(3): 1529-1555 (September 2021). DOI: 10.1214/21-AOAS1439}
\end{keyword}

\end{frontmatter}

\section{Introduction}
Cognitive diagnosis assessments provide information on an individual's latent traits -- skills, knowledge components, personality traits, or psychological disorders -- based on his or her observed responses to carefully designed items. Compared with traditional tests for measuring proficiency that often assume the existence of a unidimensional latent trait, cognitive diagnosis models (CDMs) focus on detecting the presence or absence of several distinct skills or traits, also known as latent {\it attributes}. The information on ``mastery" or ``nonmastery'' of the given skill set for a given individual -- a cognitive diagnosis -- is used for designing targeted intervention strategies to remedy those latent attributes that have not been mastered yet.

CDMs are tools that provide a cognitive diagnosis profile for every respondent. 
 Even though the earliest cognitively diagnostic models were proposed in 1980s, this topic has gained great popularity in recent years due to the advancement of computation power needed for handling complex models and also due to the models' desirable diagnostic nature of providing informative cognitive profiles for every respondent \citep{Tatsuoka,Rupp}.
Various CDMs have been developed under different assumptions 
about cognitive processes \citep[e.g.,][]{DiBello,Junker,dela,Templin,davier2008general,HensonTemplin09,dela2011}. Two notable CDMs are the basic  Deterministic Input Noisy output ``And'' gate (DINA) model  \citep{Junker}, which requires only two parameters for each item regardless of the number of attributes measured by the item, and  the more complex Generalized DINA (GDINA) model \citep[][]{dela2011}, which includes many commonly used CDMs as special cases. 
   
 Most CDMs currently assume binary classifications -- mastery or nonmastery -- for latent attributes. 
 As a consequence, when partial or incomplete  mastery is possible, heterogeneity in response data may not be well accounted for by standard CDMs. Likewise, standard CDMs with binary attributes may not be able to provide accurate classifications for subjects with immature -- yet better than guessing -- mastery skills. 
  Even though various goodness-of-fit tests have been developed in the CDM settings 
 \citep[e.g.,][]{de2008empirically,rupp2008effects,hansen2016limited,gu2018hypothesis},  to our knowledge, only  few diagnosis tools are available for testing for partial mastery under the CDM framework.  Among them, tools proposed by \cite{chen2013general} who developed a general CDM to accommodate  the expert-defined polytomous attributes, which are pre-defined as part of the test development process to provide additional diagnostic information; see also the GDM for polytomous attributes \citep{davier2008general}. 
  Moreover, due to the discreteness nature of CDMs, identifiability conditions might be difficult to satisfy in practice; for instance, it has been established that single-attribute items are necessary to identify all attribute profiles under the DINA model \citep{Chiu} and estimate the model parameters \citep{Xu15,Xu2016,gu2017sufficient}. 
  
 This paper proposes a new and flexible type of cognitive diagnosis models --  Partial-Mastery CDMs (PM-CDMs) -- that builds on the notion of multiple latent dimensions
  as in CDMs, but allows for partial mastery. 
 For each subject, we introduce a set of continuous partial mastery scores between 0 and 1 that measure his/her mastery level for each attribute; this assumption subsumes CDMs as special cases when all mastery scores are either 0 or 1. 
 This general assumption is closely related to the {\it grade of membership model} \citep{Manton,Haberman95}, also called the {\it mixed membership model}   \citep{erosheva2004mixed,erosheva2005comparing}.  
 The mixed membership model has been employed for development of statistical models for complex multivariate data in many fields, from computer science to genetics and the social and medical sciences \citep*{erosheva2002partial,airoldi2014handbook}. 
 Despite the similarity, the proposed PM-CDM class differs from the class of mixed membership models in that it assumes a (continuous) mastery score for each latent dimension (attribute) -- where the latent dimensions are pre-determined by expert knowledge about skills required for each item -- as opposed to a mixed membership among latent dimensions (attributes) in the mixed membership model that are typically unspecified but inferred from the analysis. 
In addition, the mastery score for each latent attribute in PM-CDMs is between 0 and 1, indicating the corresponding mastery level, while membership scores across different latent dimensions are  assumed to add up to one in the mixed membership models, which therefore cannot be interpreted as the mastery level of each attribute. 

To estimate the model parameters and latent mastery scores, this paper develops a Gibbs sampling algorithm  that is applicable to all PM-CDMs and further demonstrates its good performance using extensive simulation studies and two real datasets.
Moreover, under the proposed PM-CDM framework, we propose two diagnostic methods   for examining the binary assumption of the attribute mastery level. 
 We demonstrate utility of our partial mastery analysis framework on two existing real data examples from cognitive diagnosis literature --
the Fraction Subtraction and the English test data sets. The data examples show that the proposed PM-CDMs not only improve the goodness of fit over CDMs, but also could make substantial difference in conclusions about attribute mastery. Therefore, PM-CDMs would lead to more effective remediation programs in practice by providing the detailed individual-level information about the targeted ability skills.

The rest of this paper is organized as follows.   Section 2 gives a brief review of CDMs. Section 3 introduces PM-CDMs and the estimation of the model parameters. Simulation studies are further conducted for special cases of DINA and GDINA to compare the performance of the   respective PM-CDMs and CDMs in Section 4. Section 5 presents analyses of the  Fraction Subtraction and English test data sets. Section 6 provides further discussion, and the Supplementary Material \citep{supp} presents details for the proposed Gibbs sampling algorithm and additional numerical results.
  
\section{A Brief Review of CDMs}
 In cognitive diagnosis setting, a subject (e.g., examinee or patient) provides a $J$-dimensional binary response vector $\RR = (R_1,...,R_J)^\top$ to $J$ diagnostic test items, {where the superscript $\top$ denotes the transpose}.  These responses are assumed to be dependent in a certain way on $K$ unobserved latent attributes that is of practical interest.
In cognitive diagnosis, each of the $K$ latent attributes, denoted by $\alpha_k\in\{0,1\}$ for $k\in\{1,\cdots,K\},$ is assumed to be a binary indicator of the mastery  or nonmastery, respectively, of the $k$th attribute.
A complete set of $K$ latent attributes, $\aalpha= (\alpha_1,\ldots, \alpha_K)^\top$, is known as an attribute profile or a  latent subgroup/class.
Such a construction
of $\aalpha$ differs from the conventional latent class model setting and  is necessary for the diagnosis purpose. 

For instance, in educational testing, each $\alpha_k$ may represent  a certain ability or skill, and teachers may want to know whether a student  has mastered this skill or not.
For example, the Fraction Subtraction test dataset -- one of the first designed cognitive diagnostic tests in educational measurement  --
contains responses to 20 fraction subtraction problems designed to measure 8 skills that students in middle school need to learn for fraction subtraction (see the upper panel in Table \ref{attr-frac}). The original items, attributes and response data were conceived and collected by  \cite{tatsuoka1990toward}. Researchers are interested in estimating the mastery status of each of the 536 middle school students on 8 fraction subtraction attributes.
  
\begin{table}[h!]
\caption{Attributes of the Fraction Subtraction Data}
	\begin{tabular}{l l}
Attributes & Interpretation \\
\hline
$\alpha_1$ & Convert a whole number to a fraction\\
$\alpha_2$ & Separate a whole number from a fraction\\
$\alpha_3$ & Simplify before subtracting\\
$\alpha_4$ & Find a common denominator\\
$\alpha_5$ & Borrow from whole number part\\
$\alpha_6$ & Column borrow to subtract the second numerator from the first \\
$\alpha_7$ & Subtract numerators\\
$\alpha_8$ & Reduce answer to simplest form\\
\hline\\
\end{tabular}
\label{attr-frac}

{\it $Q$-matrix of the Fraction Subtraction Data}
\vspace{0.02in}

	  \begin{tabular}{c|  c c c c c c c c c c c c}
\hline
 Item & $\alpha_1$  & $\alpha_2$  & $\alpha_3$  & $\alpha_4$  & $\alpha_5$  & $\alpha_6$  & $\alpha_7$  & $\alpha_8$    \\
\hline
1 &    0 &    0 &    0 &    1 &    0 &    1 &    1 &    0    \\ 
  2 &    0 &    0 &    0 &    1 &    0 &    0 &    1 &    0  \\ 
  3 &    0 &    0 &    0 &    1 &    0 &    0 &    1 &    0 \\ 
  4 &    0 &    1 &    1 &    0 &    1 &    0 &    1 &    0  \\ 
  5 &    0 &    1 &    0 &    1 &    0 &    0 &    1 &    1   \\ 
  6 &    0 &    0 &    0 &    0 &    0 &    0 &    1 &    0   \\ 
  7 &    1 &    1 &    0 &    0 &    0 &    0 &    1 &    0  \\ 
  8 &    0 &    0 &    0 &    0 &    0 &    0 &    1 &    0  \\ 
  9 &    0 &    1 &    0 &    0 &    0 &    0 &    0 &    0   \\ 
  10 &    0 &    1 &    0 &    0 &    1 &    0 &    1 &    1 \\ 
  11 &    0 &    1 &    0 &    0 &    1 &    0 &    1 &    0 \\ 
  12 &    0 &    0 &    0 &    0 &    0 &    0 &    1 &    1   \\ 
  13 &    0 &    1 &    0 &    1 &    1 &    0 &    1 &    0   \\ 
  14 &    0 &    1 &    0 &    0 &    0 &    0 &    1 &    0   \\ 
  15 &    1 &    0 &    0 &    0 &    0 &    0 &    1 &    0  \\ 
  16 &    0 &    1 &    0 &    0 &    0 &    0 &    1 &    0  \\ 
  17 &    0 &    1 &    0 &    0 &    1 &    0 &    1 &    0   \\ 
  18 &    0 &    1 &    0 &    0 &    1 &    1 &    1 &    0  \\ 
  19 &    1 &    1 &    1 &    0 &    1 &    0 &    1 &    0   \\ 
  20 &    0 &    1 &    1 &    0 &    1 &    0 &    1 &    0  \\ 
  \hline
 \end{tabular}
 \label{Q-frac}

 \end{table}


Cognitive Diagnostic Models (CDMs) are statistical and psychometric tools that have been specifically designed 
 to estimate which skills have been mastered -- the subjects' diagnostic attribute profiles -- from their responses. CDMs achieve this by modeling the complex relationship among items, multivariate binary latent trait vector, and categorical item responses.

Many CDMs have the following two-step data generating process.
  The first step proposes a population model for the attributes profile. A common assumption is    that  the attribute profile        follows from  a population categorical distribution 
 $\aalpha \sim \mbox{Categorical}(\pp)$
  with proportion parameters 
$\pp:=(
p_{\aalpha}: \aalpha\in\{0,1\}^{K})^\top$,
where $p_{\aalpha}\in (0,1)$ and $\sum_{\aalpha\in\{0,1\}^{K}}p_{\aalpha}=1$.
The second step in CDMs follows a restricted latent class model framework with constraints that are incorporated   according to the cognitive processes.
Given a subject's attribute profile $\aaa$,   response $R_j$ to item $j$ under the corresponding model follows a Bernoulli distribution,
$ R_j  \mid \aaa  \sim \mbox{Bernoulli}(\theta_{j,\aaa})$,
where   $\theta_{j,\aaa}= P(R_j=1 \mid \aaa),$
  is the probability of   positive response to item $j$ for subjects with $\aaa$.
Parameters $\Theta = (\theta_{j,\aaa})$ are constrained by 
 the relationship between the $J$ items and the $K$ latent attributes, 
 specified through the $Q$-matrix that is   a $J\times K$ binary matrix 
with entries $q_{jk}\in\{0,1\}$ indicating the absence or presence, respectively, of a link between the $j$th item and the $k$th latent attribute.
The $j$th row vector  $\qqq_j$ of $Q$ provides the full attribute requirements for   item $j$.
For illustration, Table \ref{Q-frac} presents the $20\times 8$ binary $Q$-matrix for the Fraction Subtraction dataset.

Given an attribute profile $\aalpha$ and a $Q$-matrix $Q$, we 
write $\aaa\succeq \qqq_j  \mbox{ if }  \alpha_k \geq q_{jk} \mbox{ for any }  k \in\{1, \ldots, K\},$
and 
$\aaa\nsucceq \qqq_j  \mbox{ if there exists $k$ such that } \alpha_k < q_{jk};$
similarly we define the operations $\preceq$ and $\npreceq$.
  If $\aaa\succeq \qqq_j$, a subject with $\aaa$ has all the attributes for item $j$ specified by the $Q$-matrix and would be   ``capable'' to   answer item $j$ correctly; on the other hand, if $\aaa'\nsucceq \qqq_j$, the subject with $\aaa'$ misses some related attribute and is expected to have a smaller   response probability than subjects with $\aaa\succeq \qqq_j$.
That is, $\theta_{j,\aaa}
\geq \theta_{j,\aaa'}$  for $\aaa\succeq \qqq_j$ and  $\aaa'\nsucceq \qqq_j$; such monotonicity assumption is common to most CDMs. 
Another common assumption of CDMs is that mastering nonrequired attributes will not change the response probability, i.e.,  
 $\theta_{j,\aaa}
=\theta_{j,\aaa'}, ~\mbox{ if } \aaa\otimes \qqq_j=\aaa'\otimes \qqq_j,$
 where $\otimes$ is the element-wise multiplication operator. 
Note that this constraint implies that $\max_{\aaa\succeq\qqq_j}\theta_{j,\aaa}
=\min_{\aaa\succeq\qqq_j}\theta_{j,\aaa},$ which is a key assumption for the   identifiability of the CDM parameters \citep{Xu2016}.
 From the construction, CDMs can then be viewed as  $Q$-restricted latent class models with $2^K$ latent classes.

  CDMs with various diagnostic assumptions on the modeling of $\theta$'s have been proposed in the psychometrics literature for different application purposes. We next introduce the popularly used DINA and GDINA models, which are the representations of the basic   and general CDMs respectively. 
\paragraph{DINA model} One of the basic cognitive diagnosis model is the DINA model \citep{Junker}, which  assumes a conjunctive relationship among attributes. That is, it is necessary to possess all the attributes indicated by the $Q$-matrix to be capable of providing a correct response. In addition, having additional unnecessary attributes does not compensate for the lack of necessary attributes.
For  item $j$ and attribute vector $\aaa$, we define the ideal response
$\xi_{j,\aaa}^{DINA}= I (\aaa \succeq \qqq_j)$, i.e., the response when there is no measurement error. 
The uncertainty is further incorporated at the item level, using the slipping and guessing parameters $\sb$ and $\gg$.
 For each item $j$, the slipping parameter  $s_j = P(R_j = 0\mid \xi_{j,\aaa}^{DINA} = 1) $ denotes the probability of the respondent making a incorrect   response despite mastering all necessary skills; similarly, the guessing parameter  $g_j = P(R_j = 1\mid  \xi_{j,\aaa}^{DINA} = 0)$  denotes the probability of a correct response despite an incorrect ideal response. 
The response probability $\theta_{j,\aaa}$ then takes the form
\begin{equation}\label{DDINA}
\theta_{j,\aaa} = (1-s_j)^{\xi_{j,\aaa}^{DINA}}g_j^{1-\xi_{j,\aaa}^{DINA}}.
\end{equation}
 In practice, it is usually assumed that  $1-s_j>g_j$ for any   $j$ for model identifiability.

\paragraph{G-DINA model} 
 \cite{dela2011} generalizes the DINA model to the
G-DINA model.
The  formulation of the G-DINA model based on $\theta_{j,\aaa}$ can be decomposed into the sum of the effects due the presence of specific attributes and their interactions. Specifically, for item $j$ with $\qqq$-vector $\qqq_j=(q_{jk}: k=1, \cdots, K)$,
\begin{eqnarray}\label{GGDINA}
\theta_{j,\aaa}  &=&
\beta_{j0} 
+ \sum_{k=1}^{K}\beta_{jk}(q_{jk}\alpha_{k})
+ \sum_{k'=k+1}^{K}\sum_{k=1}^{K-1}\beta_{jkk'}(q_{jk}\alpha_{k})(q_{jk'}\alpha_{k'})\notag\\
&&+\cdots + \beta_{j12\cdots K}\prod_{k}(q_{jk}\alpha_{k}).
\end{eqnarray}
Note that  not all $\beta$'s in the above equation are included in the model. For instance, when $\qqq_j \neq \mathbf 1^\top$, we do not need parameter $ \beta_{j12\cdots K}$ since $\prod_{k}(q_{jk}\alpha_{k})=0$.
To interpret, $\beta_{j0}$ represents probability of a positive response when none of the required attributes are present;
 when $q_{jk}=1$, $\beta_{jk}$ is included in the model and it shows the change in the positive response probability as a result of mastering a single attribute $\alpha_k$; 
when $q_{jk}=q_{jk'}=1$, $\beta_{jkk'}$ is in the model and it shows the change in the positive response probability due to the interaction effect of mastery of both 
$\alpha_k$ and $\alpha_{k'}$; 
similarly, when $\qqq_j=\mathbf 1^\top$, $\beta_{j12\cdots K}$ represents the change in the positive response probability due to the interaction effect of  mastery of all the required attributes. 

\medskip  
The above model setup uses  $\RR$ and $\aalpha$ to denote the responses and attribute profile for a random subject from the population. In   a cognitive diagnostic assessment, suppose we have $N$ independent subjects, indexed by $i=1,\ldots, N$.
For the $i$th subject, we denote his/her response vector  by $ \RR_i=(R_{ij}: j=1,\ldots,J)^\top$ and his/her latent attribute profiles by $\aalpha_i = (\alpha_{ik}: k=1,\ldots,K)^\top$.  
 Under the local independence assumption, the likelihood function of the observed responses is 
 $$L_{CDM}( \pp,\Theta; \RR) = \prod_{i=1}^N\sum_{\aaa_i=\aaa\in\{0,1\}^K}\pp_\aaa  \prod_{j=1}^J \ttt_{j,\aaa}^{R_{ij}} (1-\ttt_{j,\aaa})^{1-R_{ij}}.$$
The CDM model parameters   can then be estimated by maximizing the likelihood function    via the EM-algorithm \citep[e.g.,][]{dela2009,feng2014parameter}; alternatively, researchers have proposed   Bayesian estimation approaches \citep[e.g.,][]{culpepper2015bayesian,culpepper2018improved}.  
R packages ``CDM" and ``GDINA" for CDM estimation and inference have also been recently developed by \cite{george2016r} and \cite{gdina-R}, respectively.

\section{Partial-Mastery Cognitive Diagnosis Models}
\subsection{Model setup}
We propose a partial-mastery modeling approach for cognitive diagnosis, which provides a generalization of CDMs.  
Usually, CDMs assume that each subject belongs to only one particular attribute profile $\aaa\in\{0,1\}^K$, where mastery of each skill is either $0$ or $1$.  That is, each attribute must be either fully mastered or not mastered.
Such an assumption may be too strong when certain attributes are mastered gradually, and partial mastery is possible. 
 In such a case, the usual CDMs with binary attributes may not fit the responses well and diagnostic classification of the subjects may also be affected; we will provide illustrations with our simulated and real data examples.
The proposed Partial-Mastery CDMs (PM-CDMs) relax this assumption by introducing $K$ subject-specific attribute scores $d_k\in[0,1]$, 
 which indicate the   subject's mastery level of the $k$th attribute with $d_k=1$ the highest and $d_k=0$ the lowest level.
   The attribute mastery score $\dd = (d_1,\cdots,d_K)^\top$ then contains the subject's attribute profile information.  Comparing  with CDMs in Section 2, PM-CDMs are  more flexible, with $d_k$  measuring how well a student masters the $k$th skill, 
    yet includes respective CDMs as special cases when $d_k=\alpha_k\in\{0,1\}$.

We assume the latent attribute mastery score $\dd$ follows a distribution $D(\cdot)$ on $[0,1]^K$. 
 Depending on application, different $D$ may be chosen. 
We propose to use a Gaussian copula model to allow for the dependencies among  the attributes.  
  In particular, we assume
\begin{equation}\label{m1}
\{\Phi^{-1}(d_k); k=1,\cdots,K\}^\top\sim 
N(\mu,\Sigma)
\end{equation}
where  $\Phi^{-1}$ is the inverse cumulative distribution function of a standard normal distribution, $\mu$ is a $K$-dimensional mean vector, and $\Sigma$ is a $K\times K$ covariance matrix.
An advantage of the Gaussian copula model is that it can directly characterize both the population mean mastery scores  via $\mu$ and  the latent correlations among the $K$  attributes via the underlying covariance matrix $\Sigma$.  
 Our model is similar to the correlated topic model by \cite{blei2007correlated} in the sense that it allows for modeling dependencies among the latent attributes, however, the major difference is that mastery scores for the latent attributes do not add up to one whereas topic proportions in the correlated topic model do have to add up to one. In other words, the mastery scores in PM-CDMs live on the hypercube $[0,1]^K$ while the topic proportions in the correlated topic model live on a simplex.

 Under the PM-CDM framework,  the attribute mastery score $d_k$ indicates the mastery level of the $k$th attribute and the item response function follows a similar assumption to the grade of membership models \citep{Manton,Haberman95}.
Specifically, 
 for a subject with a general mastery score   $\dd$, the marginal probability of   a positive response to item $j$ is a weighted combination of $\theta_{j,\aaa} = P(R_j=1\mid \aaa)$ with   $\theta_{j,\aaa}$   defined under the corresponding CDM:    
\begin{equation}\label{mmms}
	\ttt_{j,\dd} = P(R_j=1\mid\dd )=  \sum_{\aaa=(\alpha_1,\cdots,\alpha_K)\in\{0,1\}^K}\theta_{j,\aaa}\times p_{\aaa\mid \dd}.
\end{equation}
Here $p_{\aaa\mid \dd}=\prod_{k=1}^K{d_k}^{\alpha_k}(1-d_k)^{1-\alpha_k}\in [0,1]$, denoting the mixture weight of $\theta_{j,\aaa}$ given $\dd$.
Note that $\sum_{\aaa} p_{\aaa\mid \dd}=1$; we have $\min_{\aaa}\ttt_{j,\aaa}\leq \ttt_{j,\dd}\leq \max_{\aaa}\ttt_{j,\aaa}$
 and thus $\ttt_{j,\dd}$ are well defined.

Under the local independence assumption that responses to different items are independent given a mastery score, the probability mass function of a subject's observed responses  $\RR$ is 
\begin{eqnarray*}
P_{PMCDM}(\RR\mid \Theta,\mu,\Sigma) &=& \int_{\dd\in[0,1]^K}  
 \prod_{j=1}^J \ttt_{j,\dd}^{R_j} 
  \left(1- \ttt_{j,\dd} \right)^{1-R_j} dD_{\mu,\Sigma}(\dd).
\end{eqnarray*}

\paragraph{A restricted latent class model representation}
 From the model \eqref{mmms},  $\ttt_{j,\dd}$ can also be represented by a  restricted latent class model (RLCM). 
Consider a vector of $J\times 2^K$ latent variables $A = (\aaa^*_1,\cdots,\aaa^*_J)$, where $\aaa^*_j\in \{0,1\}^K$. 
The set of all possible $A$'s is 
${\cal A} = \prod_{j=1}^J\{0,1\}^K,$
which has $2^{K\times J}$ classes.
Define the distribution of the latent classes $A = (\aaa^*_1,\cdots,\aaa^*_J)\in {\cal A}$
as 
\begin{align}\label{piA2}
	\pi_A &= E_{D_{\mu,\Sigma}(\dd)}\prod_{j=1}^J  \prod_{k=1}^K{d_k}^{\alpha^*_{jk}}(1-d_k)^{1-\alpha^*_{jk}} \\\notag
	&= E_{D_{\mu,\Sigma}(\dd)} \prod_{k=1}^K{d_k}^{\sum_{j=1}^J\alpha^*_{jk}}(1-d_k)^{\sum_{j=1}^J(1-\alpha^*_{jk})} .
\end{align}
The conditional distribution of the responses given a latent class $A$ is modeled under the constraint that 
$P(R_j = 1\mid A)  =P(R_j = 1\mid \aaa^*_j) =\ttt_{j,\aaa^*_j}.$
The $\ttt_{j,\aaa^*_j}\in [0,1]$ is as defined in the CDMs.
Under the local independence assumption, the probability mass function of a subject's observed responses $\RR$ under the above   RLCM   is 
$$P_{RLCM}(\RR\mid \Theta,\mu,\Sigma) = \sum_{A=(\aaa^*_1,\cdots,\aaa^*_J)\in {\cal A}} \pi_A \prod_{j=1}^J \ttt_{j,\aaa^*_j}$$
Following an argument similar to that of Lemma 3.1 in \cite{Erosheva2007},  we can prove that  
$P_{RLCM}(\RR\mid  \Theta,\mu,\Sigma) = P_{PMCDM}(\RR\mid  \Theta,\mu,\Sigma).$
 Therefore, the PM-CDM is equivalent to the above RLCM with the distribution of the latent classes as specified in Equation \eqref{piA2}.

The equivalent RLCM representation plays a key role in our estimation of PM-CDM as introduced in Section 3.2. 
Moreover, with the above RLCM representation, we can study identifiability of PM-CDMs using recently developed techniques for identifiability of RLCMs \citep{Xu2016,gu2018partial}. In particular, \cite{Xu2016} and \cite{gu2018partial} studied identifiability and partial identifiability of a general family of RLCMs, including the DINA and GDINA model, and showed that these RLCMs are identifiable if the $Q$-matrix satisfies certain structural  conditions. 
With the RLCM representation,  we would expect that the identifiability of the model parameters of PM-CDMs could be established under a similar set of  structural conditions for the $Q$-matrix.

\paragraph{Relationships with CDMs} 
The key concept of PM-CDMs is to assume that  each subject has a continuous score  $d_k,$ for any $k\in\{1,\cdots, K\}$  of each attribute in the range $[0,1]$, indicating the mastery level of the $k$th attribute.
Following \eqref{mmms}, we can show that the proposed PM-CDMs cover  the   CDMs 
  as   special cases when the mastery  score   $d_k$  takes discrete  values.  
In particular, when $\dd\to \mathbf 1$, then only $p_{\mathbf 1\mid \dd}$ is significant and $\ttt_{j,\dd}\to \theta_{j,\mathbf 1}=\max_{\aaa}\ttt_{j,\aaa}$ and similarly,  when  $\dd\to \mathbf 0$, $\ttt_{j,\dd}\to \theta_{j,\mathbf 0}=\min_{\aaa}\ttt_{j,\aaa}$.
More generally, when $\dd =\aaa \in\{0,1\}^K$, i.e., the subject's mixed attribute profile is exactly a binary vector $\aaa$,  the probability of providing a positive response to  item $j$ is  $\theta_{j,\dd} = P(R_j=1\mid\dd= \aaa)= \theta_{j,\aaa}$. Therefore, with $\theta_{j,\aaa}$ defined under  any CDM, the PM-CDM covers the corresponding CDM as a special case.  

\begin{figure}[h!]
   \includegraphics[width=1\textwidth]{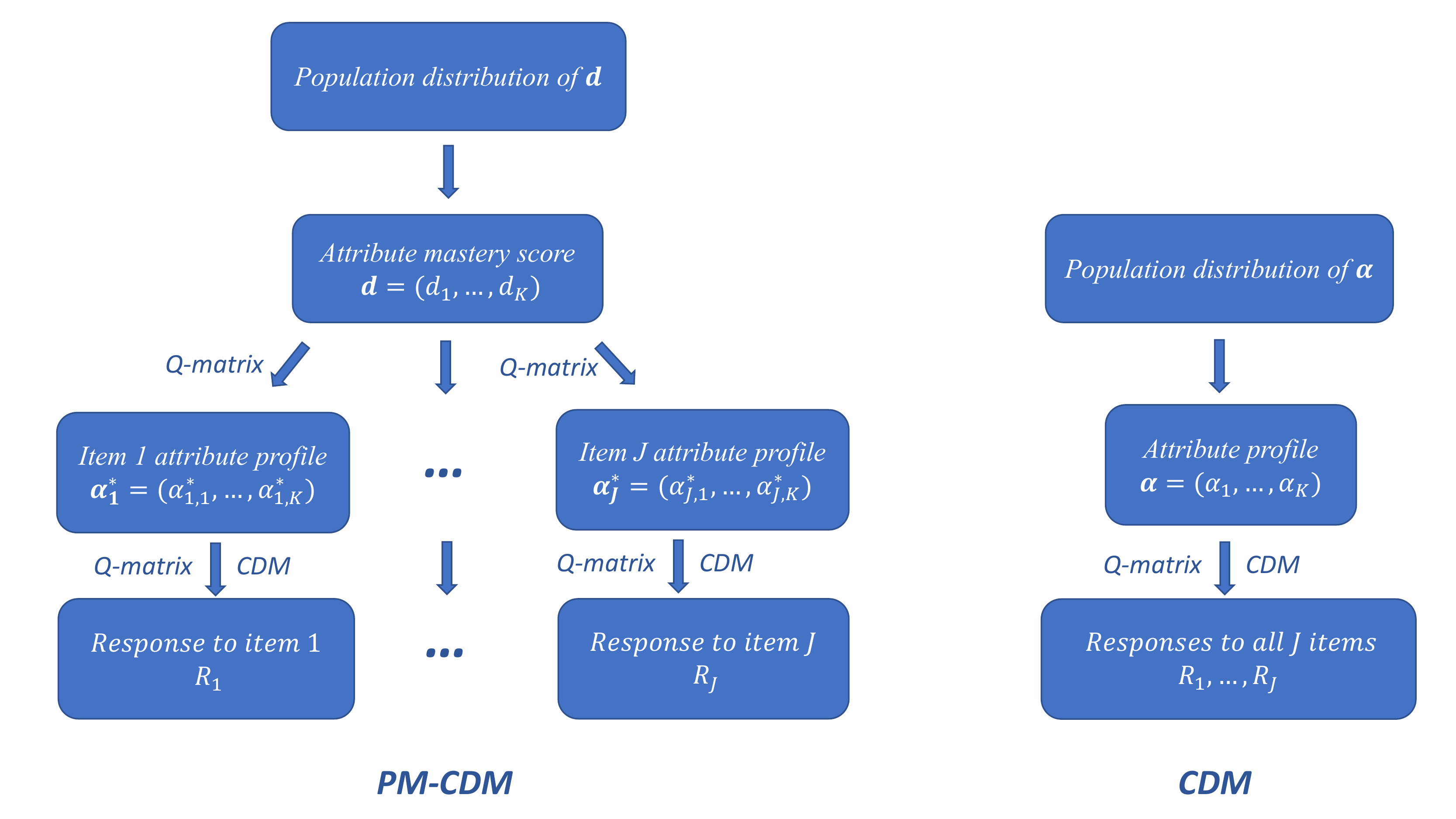}
    \caption{ {PM-CDM and CDM data generating processes for a subject}}
  \label{cdm}
  \end{figure}

 {The difference between PM-CDMs and CDMs can be further revealed under the RLCM representation of PM-CDMs.  Following Equation \eqref{piA2} and the related discussion, we  can interpret a PM-CDM corresponding to the following three-step data generating process:
 \begin{itemize}
\item[  1)]  the attribute mastery score $\dd$ is  generated from a population distribution $D_{\mu,\Sigma}(\cdot)$;
\item[  2)]  the auxiliary latent indicators for item $j$,   $\aaa_{j}^*= (\alpha^*_{jk}: k=1,\cdots,K)$, is generated such that $\alpha^*_{jk}$'s are independent across $k$ and 
$\alpha^*_{jk}\sim$ Bernoulli$(d_k)$.
\item[  3)] given the working latent attributes $\aaa_{j}^*$ for item $j$, the response $R_{j}$ is generated from  Bernoulli($\ttt_{j,\aaa_{j}^*}$). 	
\end{itemize}
The three-step data generating process is also illustrated in Figure \ref{cdm}, where for comparison, we also include an illustration of the data generating process of a CDM. 
Under this RLCM representation, we can see that the PM-CDMs allow the ``working" attribute profiles $\{\aaa_{j}^*, j=1,\ldots,J\}$ of a subject to be flexible from item to item, whereas CDMs assume the attribute profile $\aaa$ is fixed for all items. 
 Therefore, instead of performing ``hard" clustering as in CDMs, where each subject is only assigned to one latent attribute profile, the RLCM representation of PM-CDMs implies that PM-CDMs are performing ``soft" clustering of each subject, by allowing its latent attribute profile $\aaa_{j}^*$ varying from item to item. 
Note that the overall distribution of  $\aaa_{j}^*$ across all items is characterized by the partial mastery score (probability) $\dd$, which is estimated from the response data; and
as discussed above,  in the special case of $\dd =\aaa \in\{0,1\}^K$, we have $\aaa_{j}^*= \aaa$ for $j =1,\ldots, J$, and the PM-CDM reduces exactly to the corresponding CDM.   
To provide an intuitive interpretation, we can view a subject's partial mastery score $d_k$, $0<d_k<1$, as his/her mastery level of the $k$th attribute. That is, for this subject, he/she masters the $k$th attribute with a certain probability (as characterized by $d_k$) and thus the realization of $\aaa^*$  for the $k$th attribute may be 1 for some items and 0 for others. For instance, for a set of pure items measuring only this attribute with zero measurement error, the subject who only partially masters the $k$th attribute with $0<d_k<1$ may answer some items correctly and others incorrectly, which would be consistent with $\aaa^*$ for the $k$th attribute being 1 or 0 for different items.
Please also see \cite{galyardt2014interpreting} for more details on the model interpretation under the RLCM representation.
 Such flexibility of PM-CDMs provides one major feature that distinguishes them from CDMs, and that helps with accounting for additional subject-level heterogeneity in responses. This also makes the PM-CDMs distinct from the high-order CDMs, whose data generating process also follows the CDM illustration provided in Figure \ref{cdm}, though both types of models   using continuous latent traits in modeling the distribution of the discrete latent attributes. }

 We would like to emphasize that the proposed PM-CDMs are a mixed membership generalization of the binary attribute CDMs.
Under the RLCM presentation, we can see that the proposed PM-CDMs inherit a binary attribute nature from CDMs. Specifically, the PM-CDMs are constructed using the mixed membership modeling approach to capture the distribution of the binary attribute profiles $\aaa^*$  as in Steps 1) and 2), while still using the binary attribute cognitive diagnosis modeling for the item response function in Step 3. 
  In the literature, polytomous attribute CDMs \citep[e.g.,][]{chen2013general,davier2008general} have also been proposed to relax the binary assumption of the attributes.
   The PM-CDMs differ from the polytomous attribute CDMs in two major aspects: first, PM-CDMs relax the binary attribute assumption by allowing different realizations of binary attribute profiles in different items, whereas the polytomous attribute CDMs relax the assumption by  directly letting $\aaa$ to be polytomous; second, the item response function of PM-CDMs follow from the binary CDMs assumptions, whereas polytomous attribute CDMs make polytomous item response function assumptions, such as requiring polytomous entries of the $Q$-matrix in some proposed models \citep[e.g.,][]{chen2013general}.   
  Due to these differences,   the proposed PM-CDMs and polytomous CDMs may not be directly comparable.  
 On the other hand, we would like to point out that following a similar three-step data generating process, an extension from the polytomous attribute CDMs to the polytomous attribute PM-CDMs can be done by setting the $\aaa_{j}^*$ in   Step 2) to be polytomous instead of binary, and setting Step 3) to be the data generating process of the corresponding polytomous CDMs.

 Note that, as  introduced in Section 2, our focus is on CDMs 
that can be formulated as restricted latent class models. There are other CDMs not falling in the family of restricted latent class models, such as the  Reparameterized Unified Model \citep[RUM;][]{DiBello,Hartz,stout2019reparameterized}. In addition to the latent attribute $\aalpha$, the RUM also incorporates an extra probability term $P_{c_j}(\eta)$ into the $j$th item's response function with a subject-dependent continuous latent variable $\eta$ for the respondent and an additional item parameter $c_j$.
The RUM shares the similarity to the PM-CDM that they both use continuous latent variables for modeling the item response function. Nevertheless, the two models  are constructed differently. The continuous latent variable $\eta$ in RUM is introduced to account for the additional response variability after considering the binary latent attribute $\alpha$'s. The PM-CDM, on the other hand, directly models the latent attribute as a continuous mastery score, indicating the mastery level of each attribute of interest. 
In addition, the $\eta$ in RUM  is usually unidimensional, while the mastery scores $\dd$ in PM-CDM are multidimensional. 
Moreover, the PM-CDM and RUM model the continuous latent variable differently. The RUM uses a Rasch-type item response function for $P_{c_j}(\eta)$  with $\eta$  following a normal population distribution, while the PM-CDM uses a mixed membership modeling approach and the CDM-type item response functions. 
 Particularly, the introduced restricted latent class model representation of the PM-CDM makes it statistically distinct from the RUM.

\paragraph{Relationships with multidimensional item response theory (MIRT) models}

 By assuming the mastery level to be continuous, PM-CDMs also relate  to the traditional multidimensional item response theory (MIRT) models. Such continuous latent variable models have   been proposed by researchers to perform cognitive diagnosis. For instance, \cite{hong2015efficient} proposed a model  includes a noncompensatory  IRT  term and a term following the discrete attribute  DINA  model;
\cite{hartz2008fusion} proposed a  fusion model with both
IRT-based continuous latent variable and skills-based diagnosis attributes; \cite{bolt2003estimation} studied the estimation of both compensatory and noncompensatory MIRT models; and \cite{embretson2013multicomponent} presents the noncompensatory   multicomponent latent trait model for diagnosis (MLTM-D).  
 Despite the similarity, there are several differences between the proposed PM-CDMs and these existing MIRT models. 
First, the modeling construction is different. Given an individual's latent score, the item response function of PM-CDMs uses a mixture modeling framework together with cognitive diagnosis assumptions. 
 In particular, PM-CDMs can be equivalently represented by the family of the restricted latent class models mentioned above, which significantly differs from the MIRT modeling of the latent traits. 
 Therefore,   the partial mastery score is not a simple mapping of the latent trait $\theta$ in MIRT from $(-\infty,\infty)$ to $[0,1]$. 
Second, PM-CDMs have the cognitive diagnostic nature built directly into the item response function modeling, similarly to CDMs.   
Thus, PM-CDMs naturally incorporate any CDM item level cognitive diagnostic assumption, such as the conjunctive (fully nonconmpesnatiry) and disjunctive (fully compensatory) assumptions that make use of interaction terms of the attributes. 
Therefore, similarly to CDMs, PM-CDMs provide a more flexible  diagnostic modeling approach compared to MIRT. 
Third, the estimated mastery score $d_k\in[0,1]$ from PM-CDMs directly leads to a diagnostic interpretation, indicating a subject's mastery level of an attribute, while an ability trait in $(-\infty,+\infty)$ under MIRT is not directly applicable for diagnosis and certain threshold would be needed post estimation.

\subsection{PM-CDM Estimation}
To make inference on model parameters and latent variables, we propose a Gibbs sampling  Markov chain Monte Carlo (MCMC) algorithm.  We introduce the algorithm under the general PM-CDM setting   while it can be easily applied to each specific PM-CDM such as the PM-DINA and PM-GDINA models.

  From the RLCM representation in Section 3.1, 
 the $\ttt_{j,\dd}$ can  also  be interpreted from a mixture model point of view by introducing for each subject a vector of $J\times 2^K$ auxiliary  latent indicators $(\aaa^*_1,\cdots,\aaa^*_J)$, where $\aaa^*_j\in \{0,1\}^K$.
 This restricted latent class model can be equivalently represented by the    3 steps as in Figure \ref{cdm}. 
 With the auxiliary latent indicators $\aaa^*$ integrated out,    
  we can see that this model setup gives the same form of $\ttt_{j,\dd} = P(R_j=1\mid\dd )$ as in   \eqref{mmms}.
This   equivalent RLCM representation plays a key role in   designing an efficient Gibbs sampling algorithm  to estimate the model parameters as described below.

Suppose we observe $N$ independent subjects, indexed by $i=1,\ldots, N$, with their responses generated from a PM-CDM.
For the $i$th subject, we denote his/her response vector  by $ \RR_i=(R_{ij}: j=1,\ldots,J)^\top$,   latent mastery score  by $\dd_i =(d_{ik}: k=1,\ldots,K)^\top$, and    latent indicators for item $j$ by $\aalpha^*_{ij} = (\alpha^*_{ijk}: k=1,\ldots,K)^\top$.  
Denote $ \DD =\{\dd_i: i=1,\ldots, N\}$, $\AA^*= \{ \aaa^*_i: i=1,\ldots, N\}$, and $ {\cal R} = \{  \RR_i: i=1,\ldots, N\} $.  Following   the RLCM representation, the joint probability distribution of $(\DD,\AA^*, {\cal R})$ is 
 \begin{align*}
&P(\DD,\AA^*, {\cal R} \mid  \Theta,  \mu, \Sigma)\\
& =    \prod_{i=1}^N D_{\mu,\Sigma}(\dd_i)\prod_{j=1}^J\prod_{k=1}^K {d_{ik}}^{\alpha^*_{ijk}}(1-d_{ik})^{1-\alpha^*_{ijk}}
\ttt_{j,\aaa^*_{ij} }^{R_{ij}}(1-\ttt_{j,\aaa^*_{ij} })^{(1-R_{ij})}.\notag
\end{align*}
 Let $p_0(\mu)$, $p_0(\Sigma)$, $p_0(\Theta)$ be the priors of $\mu,\Sigma,$ and $\Theta$.
From the RLCM representation, we have the joint posterior distribution of $\mu, \Sigma, \Theta$: 
\begin{align}
  & P(\mu, \Sigma, \Theta, \DD,\AA^* \mid {\cal R} )
 =    ~p_0(\mu)p_0(\Sigma)p_0(\Theta)
\\
 & \times  \prod_{i=1}^N D_{\mu,\Sigma}(\dd_i)\prod_{j=1}^J\prod_{k=1}^K {d_{ik}}^{\alpha^*_{ijk}}(1-d_{ik})^{1-\alpha^*_{ijk}}
\ttt_{j,\aaa^*_{ij} }^{R_{ij}}(1-\ttt_{j,\aaa^*_{ij} })^{(1-R_{ij})}.\notag
\end{align}

 Computationally, it is more convenient to work with $\tilde \dd_i =(\tilde d_{ik}, k=1,\cdots,K)$ which is defined to be 
\begin{align} \label{tilde_d}
\tilde \dd_i:=\Phi^{-1}(\dd_i)=( \Phi^{-1}(d_{i,k}): k=1,\cdots,K).	
\end{align}
 The posterior distribution represented by $\tilde\DD :=\{\tilde \dd_i: i=1,\ldots,N\}$ is 
\begin{align} \label{post}
P(&\mu, \Sigma, \Theta, \tilde\DD,\AA^* \mid{\cal R}) 
  = p_0(\mu)p_0(\Sigma)p_0(\Theta)\prod_{i=1}^N N_{\mu,\Sigma}(\tilde\dd_i)
\\
&\times  
\prod_{j=1}^J\prod_{k=1}^K {\Phi(\tilde d_{ik})}^{\alpha^*_{ijk}}(1-\Phi(\tilde d_{ik}))^{1-\alpha^*_{ijk}} 
\ttt_{j,\aaa^*_{ij} }^{R_{ij}}(1-\ttt_{j,\aaa^*_{ij} })^{(1-R_{ij})}.\notag
\end{align}
where $N_{\mu,\Sigma}(\tilde\dd_i)$ denotes the normal distribution density with mean $\mu$ and covariance matrix $\Sigma$.

To perform the Gibbs sampling based on the posterior distribution in \eqref{post},  the difficulty arises from the sampling of  $\tilde\dd_i$. Note that the conditional distribution of $\tilde\dd_i$ is 
\begin{equation}\label{td}
	 P(\tilde\dd_i  \mid \mu, \Sigma, \Theta, \AA^*, {\cal R}) \propto    N_{\mu,\Sigma}(\tilde\dd_i)\prod_{j=1}^J\prod_{k=1}^K {\Phi(\tilde d_{ik})}^{\alpha^*_{ijk}}(1-\Phi(\tilde d_{ik}))^{1-\alpha^*_{ijk}}.
\end{equation}
 The above conditional distribution of $\tilde\dd_i$ given $\aaa^*_{ij}$'s is   nontrivial to sample directly.  To overcome this difficulty,  we use an equivalent representation of the proposed model by introducing the auxiliary variables $z_{ijk}\sim N(\tilde d_{ik},1)$ and redefining $\alpha^*_{ijk}= 1$ if $z_{ijk}\geq 0$ and  $\alpha^*_{ijk}= 0$ otherwise. We denote $\zz_i=(z_{ijk}: j=1,\ldots,J; K=1,\ldots,K)^\top$.
 With this representation, we  can first sample $z_{ijk}$ from the conditional posterior distribution   $p(z_{ijk}\mid \mu, \Sigma, \Theta, \tilde\dd_i,\AA^*, {\cal R}) \propto   
N_{\tilde d_{ik},1}(z_{ijk})I(z_{ijk}\geq 0)^{\alpha^*_{ijk}}I(z_{ijk}<0)^{1-\alpha^*_{ijk}}    $ which is a truncated Gaussian distribution with mean $d_{ik}$ and variance one; then we
  sample 
  $ \tilde\dd_i $ from Gaussian distribution  $p(\tilde \dd_{i}\mid \mu, \Sigma, \ttt,  \AA^*,\zz_i, {\cal R}) \propto   N_{\mu,\Sigma}(\tilde  \dd_{i}) \prod_{j=1}^J  \prod_{k=1}^K  
N_{\tilde d_{ik},1}(z_{ijk}).$ 

 With augmented variables ${\bf z}$ and conjugate priors, the conditional distribution of model parameters can be written in closed-form expressions. 
Sampling from the above conditional distribution of each parameter can be done   using standard statistical software such as R. 
The detailed Gibbs sampling steps are provided in the Supplementary Material \citep{supp}. 
 
\section{Simulation Studies}
There are three primary objectives of the simulation studies presented in this paper: (1) to demonstrate parameter recovery via our Gibbs sampling under the new PM-CDM approach, (2) to investigate the impact of model misspecification with respect to partial mastery, and (3) to develop diagnostic tools that could be used by practitioners to decide between CDM and PM-CDM approaches. 

\subsection{Simulation settings}

We conducted simulation studies to compare the model fit for PM-CDM and classical CDM under various data generation conditions. The proposed model is evaluated when the data are generated from either a classical CDM or a partial-mastery CDM, in terms of item parameter estimation and latent profile (attribute) specification. 
In the simulation, we consider  two popular CDMs in educational measurements, the DINA and GDINA models, and their partial mastery counterparts, PM-DINA and PM-GDINA. 
 Specifically, for each model we consider a fixed-length test of $J = 20$ and the number of latent attributes to be $K=3$ and $K=5$.
  For each $K$, we simulate two types of $Q$-matrices as given in \eqref{Q-form}: the complete -- under the DINA model -- $Q$-matrices   ($Q_3$ and $Q_5$), with an $K\times K$ identity submatrix, and the $Q$-matrices without identify submatrices ($Q'_3$ and $Q'_5$). Note that $Q'_3$ and $Q'_5$ are incomplete under the DINA model \citep{Chiu} while they are complete under the GDINA model \citep{kohn2017procedure,kohn2018build}.
 
\renewcommand{\arraystretch}{1}
\setlength{\arraycolsep}{2.8pt}
\begin{equation}\label{Q-form} 
{\scriptsize  
{\text{{$\bm Q_3$}}=\begin{pmatrix}
1 & 0 & 0 \\ 
  0 & 1 & 0 \\ 
  0 & 0 & 1 \\ 
  1 & 0 & 0 \\ 
  0 & 1 & 0 \\ 
  0 & 0 & 1 \\ 
  1 & 0 & 0 \\ 
  0 & 1 & 0 \\ 
  0 & 0 & 1 \\ 
  1 & 1 & 0 \\ 
  1 & 0 & 1 \\ 
  0 & 1 & 1 \\ 
  1 & 1 & 0 \\ 
  1 & 0 & 1 \\ 
  0 & 1 & 1 \\ 
  1 & 1 & 0 \\ 
  1 & 0 & 1 \\ 
  1 & 1 & 1 \\ 
  1 & 1 & 1 \\ 
  1 & 1 & 1 \\ 
\end{pmatrix}  
\text{{$\bm Q'_3$}}=\begin{pmatrix}
0 & 1 & 1 \\ 
  1 & 0 & 1 \\ 
  1& 1 & 0 \\ 
  0 & 1 & 1 \\ 
  1 & 0 & 1 \\ 
  1 & 1 & 0 \\ 
  0 & 1 & 1 \\ 
  1 & 0 & 1 \\ 
  1 & 1 & 0 \\ 
  1 & 1 & 0 \\ 
  1 & 0 & 1 \\ 
  0 & 1 & 1 \\ 
  1 & 1 & 0 \\ 
  1 & 0 & 1 \\ 
  0 & 1 & 1 \\ 
  1 & 1 & 0 \\ 
  1 & 0 & 1 \\ 
  1 & 1 & 1 \\ 
  1 & 1 & 1 \\ 
  1 & 1 & 1 \\ 
\end{pmatrix}   
\text{{$\bm Q_5$}}=\begin{pmatrix}
1 & 0 & 0 & 0 & 0\\ 
  0 & 1 & 0 & 0 & 0\\ 
  0 & 0 & 1 & 0 & 0\\ 
 0 & 0 & 0 & 1 & 0\\ 
  0 & 0 & 0 & 0 & 1\\ 
 1 & 0 & 0 & 0 & 0\\ 
  0 & 1 & 0 & 0 & 0\\ 
  0 & 0 & 1 & 0 & 0\\ 
 0 & 0 & 0 & 1 & 0\\ 
  0 & 0 & 0 & 0 & 1\\ 
  1 & 1 & 0 & 0 & 0\\ 
  0 & 1 & 1 & 0 & 0\\ 
  0 & 0 & 1 & 1 & 0\\ 
 0 & 0 & 0 & 1 & 1\\ 
  1 & 0 & 0 & 0 & 1\\ 
  1 & 1 & 1 & 0 & 0\\ 
  0 & 1 & 1 & 1 & 0\\ 
  0 & 0 & 1 & 1 & 1\\ 
 1 & 0 & 0 & 1 & 1\\ 
  1 & 1 & 0 & 0 & 1\\ 
\end{pmatrix}
\text{{$\bm Q'_5$}}=\begin{pmatrix}
1 & 1 & 0 & 0 & 0 \\ 
  0 & 1 & 1 & 0 & 0 \\ 
  0 & 0 & 1 & 1 & 0 \\ 
  0 & 0 & 0 & 1 & 1 \\ 
  1 & 0 & 0 & 0 & 1 \\ 
  1 & 1 & 0 & 0 & 0 \\ 
  0 & 1 & 1 & 0 & 0 \\ 
  0 & 0 & 1 & 1 & 0 \\ 
  0 & 0 & 0 & 1 & 1 \\ 
  1 & 0 & 0 & 0 & 1 \\ 
  1 & 1 & 1 & 0 & 0 \\ 
  0 & 1 & 1 & 1 & 0 \\ 
  0 & 0 & 1 & 1 & 1 \\ 
  1 & 0 & 0 & 1 & 1 \\ 
  1 & 1 & 0 & 0 & 1 \\ 
  1 & 1 & 1 & 0 & 0 \\ 
  0 & 1 & 1 & 1 & 0 \\ 
  0 & 0 & 1 & 1 & 1 \\ 
  1 & 0 & 0 & 1 & 1 \\ 
  1 & 1 & 0 & 0 & 1 \\ 
\end{pmatrix}
} }
 \end{equation}

 For the latent attributes generated under PM-CDMs,  different means and correlations are used to reflect different scenarios   of the knowledge mastery in the population.
 For $K=3$, the mean of $\tilde{\dd}$ is taken as $  \mu_3 =   {\bm 0}$ or $ (-1,0,1)^\top$, and for $K=5$, the mean is   $  \mu_5 =  {\bm 0}$ or $ (-1,-0.5,0,0.5,1)^\top$. Note that the non-constant mean represents imbalanced distribution of the latent attribute profiles.
 The covariance matrix of $\tilde{\dd}$ is 
defined  as  $ \Sigma = \sigma^2 \{\rho \mathbf{1} \mathbf{1}^T + (1-\rho) I_K\}$, where $\mathbf{1} = (1,..,1)^\top$ and $I_K$ is the $K\times K$ identity matrix. The diagonal terms of covariance matrix are fixed to be $\sigma^2 = 1$ and the correlation parameter $\rho$ is taken as $0$ or $0.8$, corresponding to low and high correlations of the latent attributes. 
Since CDMs can be considered as special cases of PM-CDMs with each $d_k\in\{0,1\}$, when we simulate data from the CDMs, we follow the PM-CDM data generating process by first simulating $\dd$'s  and then round them to be 0 or 1.  
Note that the used method of simulating correlated attributes under CDMs are the same to many existing studies \citep[e.g.,][]{Chiu,Chen2014}

 After generating latent attributes $\bm \tilde{d}_i$ from $  \mbox{Normal}(  \mu,   \Sigma)$ for $i =1,..,N$, we generate  data under conventional CDMs and PM-CDMs. In particular,  datasets are simulated  according to 4 models:   DINA, PM-DINA, GDINA and PM-GDINA. The correct response probability for DINA and PM-DINA model is set to be $0.2$ (non-mastery) and $0.8$ (mastery); in other words, both the slipping and guessing parameters are set to be $0.2$ for all items.
 For GDINA and PM-GDINA, there is an additive effect of the mastered attributes on the probability. For items requiring only 1 attribute, it is either $0.2$ (non-mastery) or $0.8$ (mastery);  for 2-attribute items, it is $0.2$ (master none), $0.5$ (master 1 of 2) or $0.8$ (master both); for 3-attribute items, it's $0.2$ (master none), $0.4$ (master 1 of 3),  $0.6$ (master 2 of 3) or $0.8$ (master all). 

We consider two sample sizes for each $K$. For $K=3$, we generate $N=500$ and $N = 1000$ subjects; for $K=5$ we use $N=1000$ and $N = 2000$. 
In summary, this simulation study uses the following 128 conditions:
$$
  \begin{pmatrix}
\hbox{DINA} \\
\hbox{PM-DINA}\\
\hbox{GDINA}\\
\hbox{PM-GDINA}
 \end{pmatrix} 
 \otimes 
 \begin{pmatrix}
K = 3 \\
K = 5
 \end{pmatrix} 
 \otimes 
 \begin{pmatrix}
Q_K \\
Q_K'
 \end{pmatrix} 
 \otimes 
\begin{pmatrix}
 \mu =   {\bm 0}  \\
  \mu \neq  {\bm 0} 
 \end{pmatrix}
 \otimes 
\begin{pmatrix}
\rho = 0 \\
\rho = 0.8
 \end{pmatrix}
 \otimes 
 N's  
 .
 $$
 For each condition, we generate 50 replications. 
 
To get a direct comparison to CDMs, we fit  both the DINA and   PM-DINA models for   the DINA- and PM-DINA-generated datasets, and both the G-DINA and PM-GDINA models   for the the GDINA- and GPM-DINA-generated datasets.
For PM-CDM estimation, we perform  the proposed Gibbs sampling  algorithm. In particular, we   take weakly-informative priors of $\mu,\Sigma$ and $\theta$'s: for $\mu$, we take  $\mu_0 = 0$ and $\Sigma_0 = I_K$; for $\Sigma$, we take $\Psi_0 = I_K$ and  $\nu_0=K+1$; for $\theta$'s,  we take weakly-informative priors $\text{Beta}(a_0=1,b_0=2)$ for $\aaa^*=\mathbf 0$ and $\text{Beta}(a_0=2, b_0=1)$ for $\aaa^*\succeq \bm q_j $, and   noninformative priors $\text{Beta}(a_0=1, b_0=1)$ for other $\aaa^*$'s.
Such Beta priors are set to reflect the assumption  that students mastering no skills will have lower probabilities to get the correct response and those   mastering all required skills of item $j$ will have higher chances to answer it correctly \citep[e.g.,][]{chen2018bayesian}. 
We take the following  initial values:
 $\mu^{(0)} = (0,...,0)^T; \Sigma^{(0)}  = E; $ 
 $ \boldsymbol d^{(0)} \sim \text{Unif}(0.01,0.99); \tilde \dd^{(0)} = \Phi^{-1}(d^{(0)})$; 
 $
z_{ijk}^{(0)} \sim \mbox{Normal}(\tilde \dd^{(0)}_{ik},1); \aaa^{(0)}_{ij} = \mathbf{1}(\boldsymbol z_{ij}^{(0)} \geq 0);
 $ 
 $
\ttt^{(0)}_{jl} = 0.5 \text{ ; } \forall j = 1,..,J; l= 1,...,2^K
 $.

We run MCMC chains in R with a total number of $M = 30,000$    and burn-in size at $10,000$  in the simulation studies. 
 Using Macbook Pro with 3 GHz Dual-Core Intel Core i7 CPU,  the computational time of the proposed Gibbs sampling algorithm is about 300 seconds for 1000 MCMC iterations under the setting of $K=3$ and $N=500$ and about 690 seconds for 1000 MCMC iterations under the setting of $K=5$ and $N=1000$, meanwhile the Bayesian DINA takes 100 and 230 seconds, respectively. As $K$ increases, the proposed algorithm gradually gains advantage over the Bayesian CDM. Under a case of $K = 10$, $N = 1000$, the estimation of the PM-CDM takes about 45 minutes whereas classical Bayesian CDM would take 95 minutes.
 Overall, as $N$ increases, the computational time increases approximately linearly in $N$. Though $M = 30,000$ and burn-in size  $10,000$  are used in our simulation, we find that the MCMC chain converges well after about 1000 iterations (the convergence of the MCMC chains is diagnosed with trace plots and Gelman-Rubin's $R^2$ $<1.1$); therefore, a few thousands iterations would suffice in practice. 
 We note that the EM algorithm for CDM estimation is computationally more efficient (taking about 5 minutes for the case of $K=10$); nevertheless, the MCMC estimation provides the full distributional information of the estimates, and its computational time may not be directly comparable to that of the EM algorithm, which usually only provides the point estimation (and standard error) results. 
 It would be an interesting future research direction to study a more efficient EM-type estimation algorithm for the proposed models.

\subsection{Simulation Results}
\subsubsection{Item Parameter Recovery Results} 
For each simulated dataset, we first compare  the CDMs and PM-CDMs in terms of their estimation of item parameters. 
Table  \ref{K3Qcom} summarizes  the simulation results for   $(K=3, N=500, Q=Q_3)$ and $(K=5, N=1000, Q=Q_5)$, respectively.  Results for other simulation settings  are presented in the Supplementary Material \citep{supp} due to space constraint. 
For the estimation of item parameters, we report the mean absolute errors (MAEs) of the posterior mean estimates of the item parameters ($\theta_{j,\aaa}$'s),  averaged over 50 replications, and the corresponding  averaged root mean squared errors (RMSEs), which are in the columns ``MAE" and ``RMSE"  of Table \ref{K3Qcom}.
When the  PM-DINA and PM-GDINA  are the true models, simulation results show that the proposed Gibbs sampling algorithm estimates the item parameters accurately with small MAEs and RMSEs. Nonconstant means and high correlations of the mastery scores have little effect on estimating the items parameters, compared with the cases with constant mean and zero correlation.    

\paragraph{Model misspecification impact} When the DINA is fitted for these data generated from the PM-DINA, the DINA model   has severe misfitting and much larger MAEs and RMSEs for the item parameters.  Similarly, we observe that the PM-GDINA outperform the GDINA in terms of item parameter estimation for all simulation  cases when the PM-GDINA is the true model.
 On the other hand,  when DINA or GDINA is the true model, they
 perform better than the PM-DINA or PM-GDINA, respectively, though the differences are minor   especially for the GDINA model with larger $K=5$, lower sample size, and more complicated settings with nonzero means and high correlations. 
This indicates that performance of PM-DINA or PM-GDINA in item parameter estimation is comparable to that of DINA or GDINA, due to the fact that CDMs can be viewed as a submodel of PM-DCMs with extremal $\dd$ values.

\begin{table}[h!]
\caption{ MAE is the mean absolute error of item parameter estimates; RMSE is the root mean squared error of item parameter estimates; AMCR is the attribute level misclassfication rate; ARSE is the root mean squared error of  mastery score   estimates.}
{\it Simulation results for $K=3$, $Q=Q_3$, and $N=500$.}
\label{K3Qcom}\centering
\resizebox{1 \textwidth}{!}{
\begin{tabular}{rrr | rrrr |  rrr}
  \hline    \multicolumn{3}{c|}{}  & \multicolumn{7}{c}{True Model}        \\  
 \multicolumn{3}{c| }{ }  & \multicolumn{4}{c}{PM-DINA}  & \multicolumn{3}{c }{ DINA} \\ 
 \hline
$\mu$ & $\rho$  & Fitted Model & MAE & RMSE & AMCR & ARSE & MAE & RMSE & AMCR  \\ 

\hline
\multirow{4}{*}{Constant} & \multirow{2}{*}{0} & PM-DINA   & 0.051 & 0.065 & 0.281 & 0.235 & 0.058 & 0.071 & 0.069 \\ 
   &  &  DINA  & 
   0.154 & 0.182 & 0.290 & 0.257 & 0.026 & 0.034 & 0.068
    \\ 
    & \multirow{2}{*}{0.8} & PM-DINA & 0.044 & 0.057 & 0.207 & 0.189  & 0.040 & 0.050 & 0.061 \\ 
   &  &  DINA & 
    0.127 & 0.145 & 0.217 & 0.223 & 0.022 & 0.028 & 0.050 \\
    \hline
  \multirow{4}{*}{Non-const.} & \multirow{2}{*}{0}  & PM-DINA & 0.053 & 0.071 & 0.194 & 0.215 & 0.066 & 0.080 & 0.056  \\ 
   &  &  DINA 
    & 0.161 & 0.205 & 0.242 & 0.243 & 0.033 & 0.047 & 0.055\\
     & \multirow{2}{*}{0.8}  & PM-DINA & 0.045 & 0.061 & 0.175 & 0.191 & 0.051 & 0.063 & 0.043  \\ 
   &  &  DINA 
     & 0.149 & 0.189 & 0.250 & 0.253 & 0.029 & 0.039 & 0.041 \\
    \hline     \multicolumn{3}{c|}{}  & \multicolumn{7}{c}{True Model}        \\  
 \multicolumn{3}{c| }{ } & \multicolumn{4}{c }{PM-GDINA} & \multicolumn{3}{c}{ GDINA}   \\ 
 \hline
$\mu$ & $\rho$ & Fitted Model & MAE & RMSE & AMCR & ARSE  & MAE & RMSE & AMCR \\  
  \hline
\multirow{4}{*}{Constant} & \multirow{2}{*}{0} & PM-GDINA  & 0.058 & 0.073 & 0.285 & 0.238& 0.076 & 0.090 & 0.084 \\ 
   &  &  GDINA 
    & 0.125 & 0.144 & 0.308 & 0.273 & 0.043 & 0.054 & 0.094 \\
     & \multirow{2}{*}{0.8} & PM-GDINA & 0.059 & 0.072 & 0.207 & 0.185 & 0.067 & 0.083 & 0.053 \\ 
   &  &  GDINA 
   & 0.092 & 0.104 &   0.249 & 0.234  & 0.057 & 0.076 & 0.084\\
        \hline
  \multirow{4}{*}{Non-const.} & \multirow{2}{*}{0} & PM-GDINA & 0.065 & 0.081 & 0.196 & 0.229 & 0.072 & 0.087 & 0.067 \\ 
   &  &  GDINA 
    & 0.114 & 0.133 & 0.328 & 0.332 & 0.053 & 0.071 & 0.061 \\ 
     & \multirow{2}{*}{0.8} & PM-GDINA & 0.060 & 0.075 & 0.161 & 0.186 & 0.068 & 0.081 & 0.051 \\ 
   &  &  GDINA
   & 0.093 & 0.118 & 0.218 & 0.240  & 0.063 & 0.087 & 0.046 \\
     \hline \\
\end{tabular}
}
{\it Simulation results for $K=5$, $Q=Q_5$, and $N=1000$}.
\vspace{0.02in}
 \label{K5Qcom}
\centering
\resizebox{1\textwidth}{!}{
\begin{tabular}{rrr | rrrr | rrr }
  \hline
   \multicolumn{3}{c|}{}  & \multicolumn{7}{c}{True Model}        \\  
 \multicolumn{3}{c| }{}  & \multicolumn{4}{c}{PM-DINA} & \multicolumn{3}{c }{ DINA}  \\ 
 \hline
$\mu$ & $\rho$ & Fitted Model & MAE & RMSE & AMCR & ARSE & MAE & RMSE & AMCR  \\ 
  \hline
\multirow{4}{*}{Constant} & \multirow{2}{*}{0}  & PM-DINA  & 0.054 & 0.070 & 0.328 & 0.255& 0.077 & 0.089 & 0.126 \\ 
   &  & DINA 
    & 0.150 & 0.178 & 0.328 & 0.262  & 0.021 & 0.028 & 0.125 \\
     & \multirow{2}{*}{0.8}  & PM-DINA & 0.040 & 0.052 & 0.217 & 0.198 & 0.052 & 0.061 & 0.073 \\ 
   &  & DINA  
   & 0.119 & 0.137 & 0.289 & 0.279 & 0.026 & 0.020 & 0.069 \\
        \hline
  \multirow{4}{*}{Non-const.} & \multirow{2}{*}{0}  & PM-DINA & 0.056 & 0.075 & 0.239 & 0.239 & 0.077 & 0.088 & 0.098 \\ 
   &  & DINA 
   & 0.158 & 0.201 & 0.316 & 0.285  & 0.034 & 0.052 & 0.090\\ 
     & \multirow{2}{*}{0.8}  & PM-DINA & 0.042 & 0.055 & 0.181 & 0.192 & 0.061 & 0.075 & 0.067 \\ 
   &  & DINA 
    & 0.156 & 0.196 & 0.290 & 0.261 & 0.044 & 0.061 & 0.064 \\
     \hline  
   \multicolumn{3}{c|}{}  & \multicolumn{7}{c}{True Model}        \\  
  \multicolumn{3}{c| }{} & \multicolumn{4}{c}{PM-GDINA}  & \multicolumn{3}{c }{ GDINA}   \\ 
 \hline
$\mu$ & $\rho$ & Fitted Model & MAE & RMSE & AMCR & ARSE & MAE & RMSE & AMCR  \\ 
  \hline
\multirow{4}{*}{Constant} & \multirow{2}{*}{0} & PM-GDINA & 0.064 & 0.078 & 0.324 & 0.256 & 0.088 & 0.101 & 0.134 \\ 
 &  & GDINA  
& 0.114 & 0.135 & 0.376 & 0.367  & 0.045 & 0.058 & 0.135 \\ 
     & \multirow{2}{*}{0.8}  & PM-GDINA & 0.069 & 0.085 & 0.215 & 0.196 & 0.070 & 0.082 & 0.082 \\ 
&  & GDINA 
 & 0.105 & 0.112 & 0.317 & 0.284 & 0.067 & 0.078 & 0.078\\
        \hline
  \multirow{4}{*}{Non-const.} & \multirow{2}{*}{0}  & PM-GDINA & 0.072 & 0.089 & 0.253 & 0.252 & 0.075 & 0.098 & 0.111 \\ 
   &  & GDINA  
    & 0.133 & 0.162 & 0.319 & 0.317 & 0.056 & 0.069 & 0.108\\ 
     & \multirow{2}{*}{0.8} & PM-GDINA & 0.071 & 0.089 & 0.195 & 0.203 & 0.072 & 0.087 & 0.078 \\ 
     &  & GDINA  
    & 0.118 & 0.133 & 0.325 & 0.312  & 0.071 & 0.085 & 0.074 \\ 
     \hline
\end{tabular}
}

\end{table}

\subsubsection{Attribute Mastery Recovery  Results} 
We approach  evaluating the estimation of  attribute profiles or mastery scores in different ways that depend on the true assumption about subject-level heterogeneity.  
We consider two cases: (i) If the true model assumes Binary mastery (CDMs), the models are evaluated by mis-classification rates. In particular,  we consider the Attribute-level Mis-Classification Rate (AMCR) that is the proportion of incorrectly classified attributes among all subjects and all attributes, i.e., 
 $\mbox{AMCR} = \sum_{i=1}^{N}\sum_{k=1}^{K} |\alpha_{ik} - \hat{\alpha}_{ik}|/{NK},$
 where $\hat{\alpha}_{ik}\in \{ 0, 1\}$ is the estimated   $k$th latent attribute for the $i$th individual: when CDMs are fitted to the data, it is estimated based on the maximum likelihood estimator and  when PM-CDMs are fitted to the data, the estimated $\hat{\alpha}_{ik}$ is computed by rounding the estimated latent mastery score  $d_{ik}$  to the nearest binary class. 
(ii) When the true model assumes partial mastery (PM-CDMs), we evaluate models by the root mean of squared error in estimating the latent attribute:  
$\mbox{ARSE} =(\sum_{i=1}^{N}\sum_{k=1}^{K} (d_{ik} - \hat{d}_{ik})^2/{NK})^{1/2},$
where  $\hat{d}_{ik}$ is the estimated latent mastery score: when CDMs are fitted to the data,  we take the posterior probabilities of latent attributes $\hat{P}(\alpha_{ik} = 1)$ as $\hat{d}_{ik}$, and  when PM-CDMs are fitted to the data, $\hat{d}_{ik}$ is   the posterior mean estimator from the MCMC algorithm. 
In addition, to have a direct comparison of the CDMs and PM-CDMs, we also compute  AMCR, where the ``true"   attribute profile of the $i$th individual  is set to be $I(\dd_i \geq 0.5)$; then for PM-CDMs we use $I(\hat{\dd}_{i} \geq 0.5)$ as the estimator, and for CDMs we use the maximal likelihood estimators.

  Table \ref{K3Qcom}  summarizes simulation results for $(K=3, N=500, Q=Q_3)$ and $(K=5, N=1000, Q=Q_5)$, respectively, while   results for other settings are reported in the Supplementary Material \citep{supp}. 
 We can see that  the PM-CDMs outperform the CDMs in most cases. 
In particular, when the PM-DINA is the true model, the DINA model has  higher  AMCR for latent attributes than the PM-DINA. On the other hand, when the DINA is the true model, the latent attribute classification results from the PM-DINA are   comparable to the true  DINA model.
   Similarly, we observe that the PM-GDINA outperform the GDINA in all cases when the PM-GDINA is the true model, while being comparable to the GDINA model when the GDINA is true. 
  
\subsubsection{Model Diagnosis of CDMs Using  Mastery Score Plots}
In this subsection, we propose two types of diagnosis for the binary assumption of the attribute mastery status under CDMs and use simulation results to illustrate their performance.

The first diagnosis method is to examine the estimated covariance matrix from PM-CDMs.   
 From the definition of PM-CDMs,  the covariance matrix $\Sigma$ in \eqref{m1} characterizes the dependent structure of the latent scores $d_k\in [0,1], k=1,\ldots, K$, up to the $\Phi^{-1}$ transformation. Since CDMs can be viewed as special cases of PM-CDMs with $d_k\in\{0,1\}$, under the CDMs, the corresponding ${\tilde d}_k=\Phi^{-1}(d_k)$ as defined in \eqref{m1}   would diverge to $\{-\infty, +\infty\}$. Therefore,  we would expect that when the data are generated from the CDMs, the diagonal elements of the estimated covariance matrix  $\hat \Sigma$  of the transformed variables $\tilde d$'s from the PM-CDMs would be relatively larger than that in the case when the data are generated from the PM-CDMs with $d_k\in [0,1]$. 
Thus, examination of diagonal elements of  $\hat \Sigma$ fitted by the PM-CDMs can then be used as an exploratory tool for the diagnosis of the binary assumption of the attribute mastery status under the CDMs. 
 We have done a simulation study to confirm our observation above. 
Table \ref{K3Var} reports covariance estimation results from the PM-CDMs under   simulation settings in Table  \ref{K3Qcom}, when the data are generated either from the CDMs or PM-CDMs. We can observe that when  the PM-CDMs are  true,  the estimated diagonal terms are close to the true value and the estimated correlation $\hat \rho$ is close to the truth; on the other hand,  when the CDMs are true, the diagonal terms $\hat{\sigma}^2$ estimated from the PM-CDMs   tend to be large. Based on these results, in practice,  when some of the estimated diagonal elements of $\hat{\bm \Sigma}$'s is not large, 
it may imply that  the CDM's assumption on the corresponding attribute  does not   hold; see real data examples in Section 5. 

%
\begin{table}[]
\caption{Covariance estimation results for $K=3$, $Q=Q_3$ and $N=500$. The fitted models are PM-CDMs and values in parentheses are the standard errors. }
\label{K3Var}
\centering
\resizebox{1  \textwidth}{!}{
\begin{tabular}{rrr | rrr | c}
 \hline
$\mu$ & $\rho$ & True Model &\multicolumn{3}{c| }{$\hat{\sigma}^2$} & $\hat{\rho}$ \\ 
  \hline
\multirow{4}{*}{Constant} & \multirow{2}{*}{0} & P-DINA & 1.03(0.43) & 1.02(0.48) & 0.99(0.42) & -0.01(0.15) \\ 
   &  & DINA & 11.84(3.64) & 11.77(3.49) & 11.07(3.28) & -0.01(0.07) \\ 
    & \multirow{2}{*}{0.8}  & P-DINA & 1.30(0.48) & 1.31(0.49) & 1.52(0.91) & 0.67(0.06) \\ 
   &  & DINA & 19.08(6.18) & 16.22(3.73) & 14.85(5.20) & 0.70(0.05) \\ 
    \hline
  \multirow{4}{*}{Non-const.} & \multirow{2}{*}{0}  & P-DINA & 1.00(0.37) & 1.14(0.59) & 0.99(0.25) & 0.02(0.16) \\ 
   &  & DINA & 7.50(1.68) & 9.87(3.69) & 4.25(1.37) & 0.03(0.11) \\ 
    & \multirow{2}{*}{0.8} & P-DINA & 1.01(0.29) & 1.50(0.93) & 1.64(0.61) & 0.64(0.08) \\ 
  &  & DINA & 10.84(2.79) & 11.53(3.24) & 4.00(1.19) & 0.65(0.14) \\ 
   \hline
   \multirow{4}{*}{Constant} & \multirow{2}{*}{0} & P-GDINA & 1.16(0.88) & 1.04(0.32) & 0.87(0.20) & 0.05(0.12) \\ 
   &  & GDINA & 4.40(1.28) & 4.51(1.58) & 4.54(2.40) & 0.03(0.07) \\ 
    & \multirow{2}{*}{0.8} & P-GDINA & 1.21(0.45) & 1.19(0.42) & 1.42(0.53) & 0.68(0.07) \\ 
   &  & GDINA & 9.91(4.41) & 9.60(4.21) & 9.90(5.30) & 0.73(0.05) \\ 
    \hline
  \multirow{4}{*}{Non-const.} & \multirow{2}{*}{0}  & P-GDINA & 0.78(0.34) & 1.08(0.63) & 0.94(0.33) & 0.04(0.14) \\ 
   &  & GDINA & 3.04(0.89) & 4.65(2.05) & 3.56(1.23) & 0.01(0.09) \\ 
    & \multirow{2}{*}{0.8} & P-GDINA & 1.24(0.88) & 1.46(0.49) & 1.36(1.16) & 0.62(0.09) \\ 
   &  & GDINA& 4.06(1.16) & 10.49(5.34) & 4.33(1.25) & 0.64(0.12) \\ 
       \hline
\end{tabular}
}
\end{table}

  \begin{figure}[h!]
\centering

\begin{minipage}{1\textwidth}
 \centering
 \includegraphics[width=0.31\textwidth]{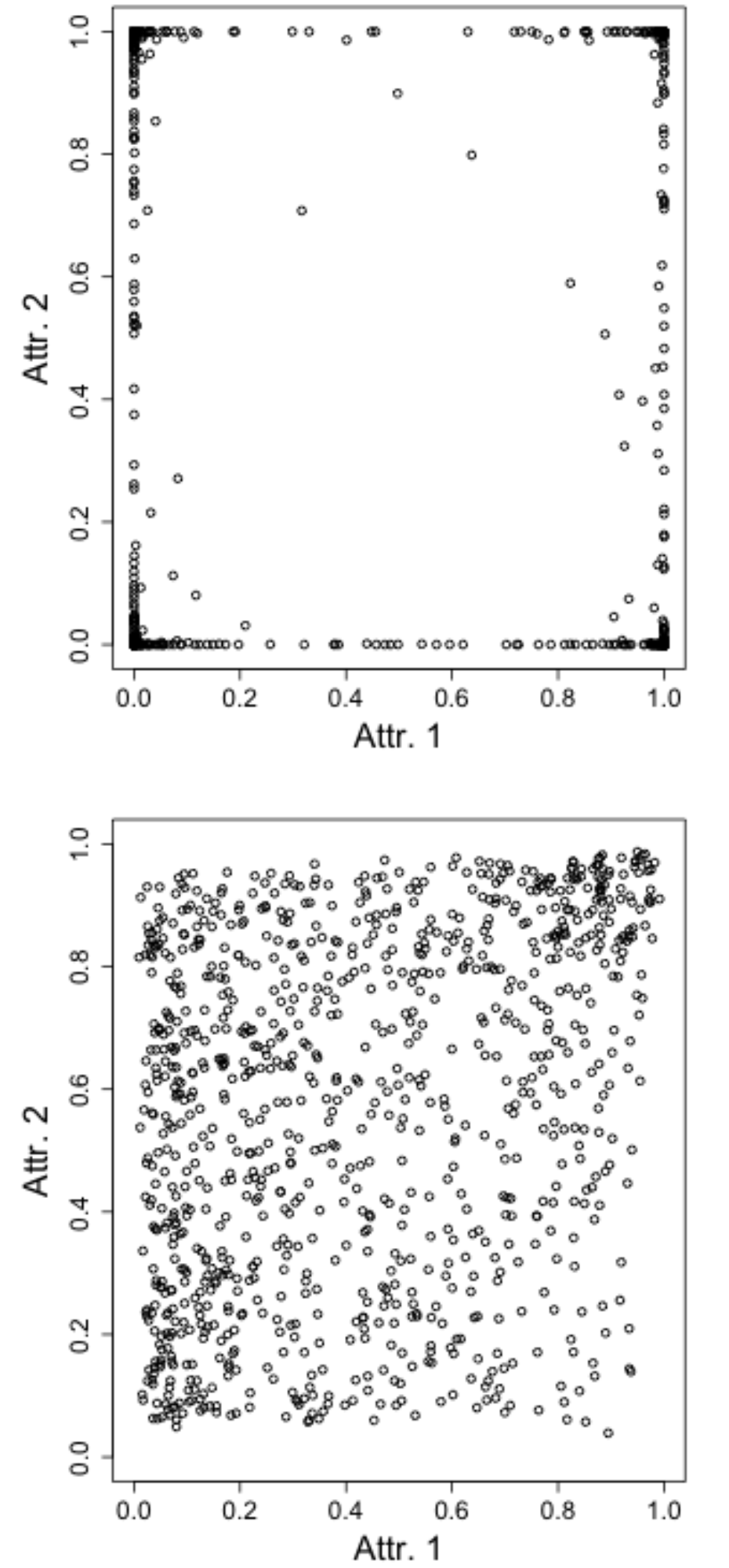}
  \includegraphics[width=0.31\textwidth]{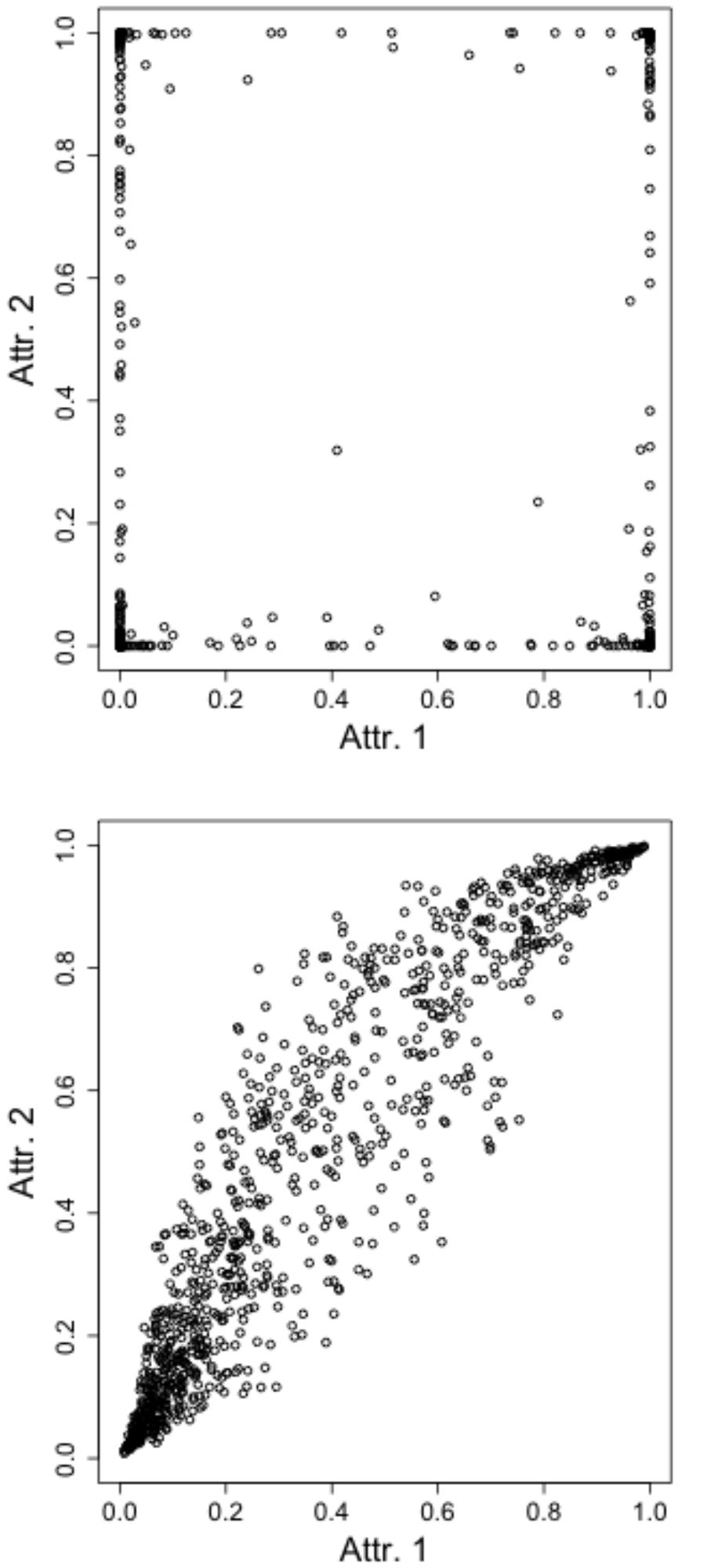}
  \includegraphics[width=0.31\textwidth]{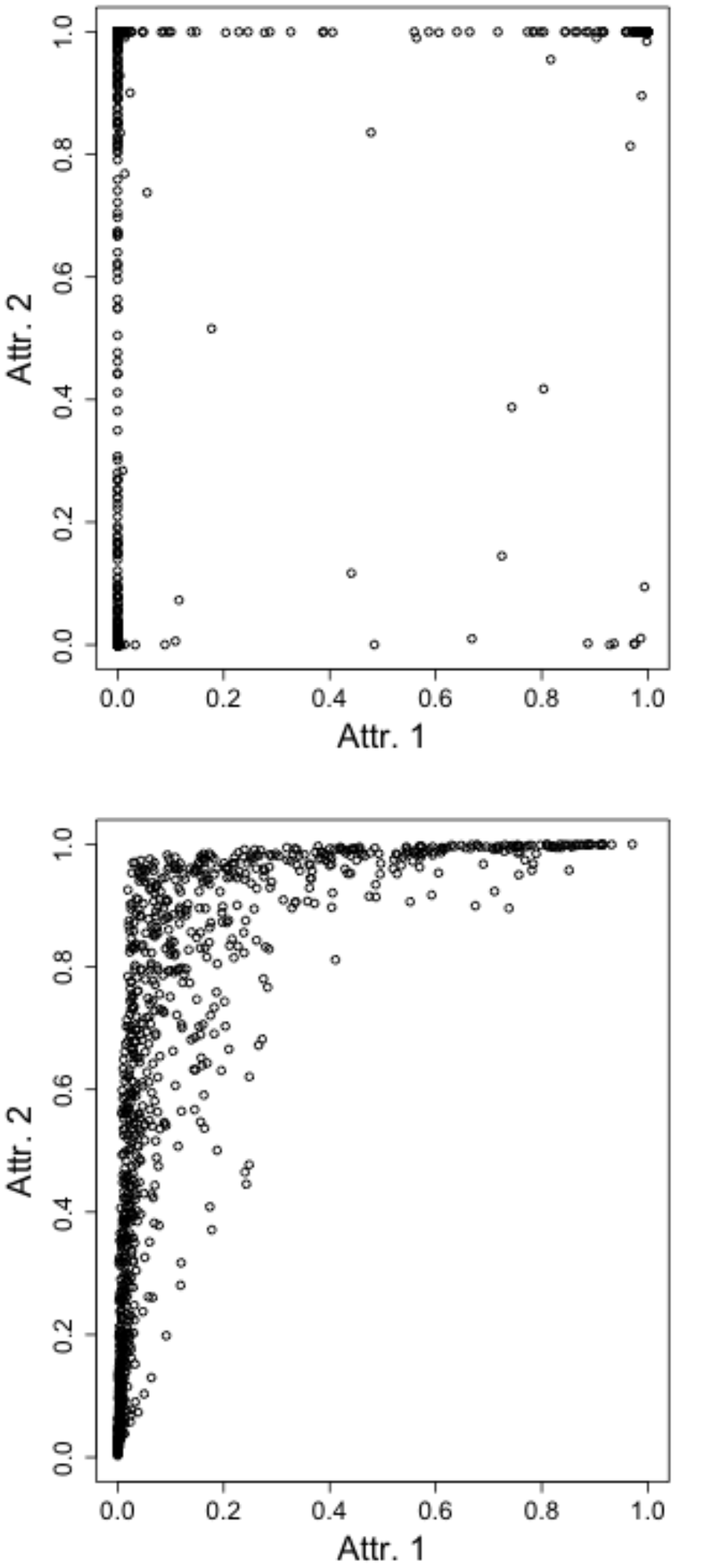}
\end{minipage}
 \caption{Estimated d of the three attributes by PM-DINA when the true model is DINA (Row 1) or PM-DINA (Row 2), under three different settings with $\mu = (0, 0, 0)$  and $\rho=0$ (Column 1),  $\mu = (0, 0, 0)$  and $\rho=0.8$ (Column 2), $\mu = (-1, 0, 1)$ and $\rho=0.8$ (Column~3). }
 \label{fig3}
\end{figure}

%
%
%
%
%

The second diagnosis method  is to examine the scatter plot of the latent attribute mastery scores estimated from  PM-CDMs. 
 If  binary mastery is true, the estimated mastery scores $\hat d$'s from PM-CDMs would be close to 0 or 1; on the other hand, if   partial mastery is true, we would expect $\hat d$'s are distributed between 0 and 1. 
To illustrate this, Figure  \ref{fig3}   report the scatter plots for the settings in the upper panel of Table \ref{K3Qcom} (the data are simulated from DINA or PM-DINA with $K=3$); simulation results of other settings are presented in the Supplementary Material \citep{supp}. 
From Figure  \ref{fig3}, when the DINA is the true model, we can see that the estimated   attribute mastery scores are concentrated at the corners, corresponding to the binary assumption of the mastery status of the attributes. 
 On the other hand, if the true model are the PM-DINAs (with $\sigma=1$), we can observe many data points spread in the middle part of the plot. 
 Therefore, in practice, if the plot of the   attribute mastery scores shows that most points are on the edges or corners, it may be preferable to use CDMs; on the other hand, if   most points are concentrated in the middle, PM-CDMs may be a better fit.

\section{Real Data Examples}\label{datasec}
\subsection{Fraction Subtraction Data}
We first consider  the Fraction Subtraction Data discussed in Table \ref{attr-frac}, which is a dataset containing responses from 536 middle school students to 20 fraction subtraction problems. 
   Based on the understanding of the problem solving process,   eight latent attributes required to answer the items   and the $20\times  8$ $Q$-matrix were specified \citep[e.g.][]{tatsuoka1990toward,dela}, which are   given in Table \ref{Q-frac}. 
 The data have been analyzed by many researchers, including \cite{TatsuokaC},  \cite{dela},   \cite{Xu2017}, among others.  
 Identifiability issue of the DINA model for this dataset has also been discussed in \cite{deCarlo2011}. 
 \cite{gu2018partial} recently proved that under the given $Q$-matrix, the DINA item parameters are all identifiable and the proportion parameters are partially identifiable.  
 
 With the $Q$-matrix, DINA, GDINA, high-order DINA (HO-DINA), high-order GDINA (HO-GDINA), PM-DINA and PM-GDINA are fitted.  The information criteria values in Table 7 in the Supplementary Material \citep{supp} show that the PM-DINA is a much better fit to these data  comparing to the other models.
 The difference in BIC/AIC values provides a very strong evidence for supporting this conclusion, according to guidelines suggested by \cite{kass1995bayes}.
 To compare the DINA and PM-DINA models, the estimated correct response probability of   non-mastery subjects ($\theta_0$, the guessing parameter) and mastery subjects ($\theta_1$, 1 $-$ the slipping parameter) are also reported in Table \ref{tbl:frac7}. 
 We can see that the PM-CDMs generally give  smaller slipping parameters on all items and smaller guessing parameters on all but one item (item 3), which agrees with the simulation estimation results when the data are generated from the PM-DINA model.

\begin{table}[h!]
\caption{Q-matrix for Fraction Subtraction data and the estimation results.}\centering
\bgroup
\def\arraystretch{1}
 \begin{tabular}{c  c c c c c c c c c c c c}
\hline
Item & \multicolumn{8}{c}{Q} & \multicolumn{2}{c}{PM-DINA}  & \multicolumn{2}{c}{DINA}\\
& A1 & A2 & A3 & A4 & A5 & A6 & A7 & A8 & $\hat{\theta}_0$ & $\hat{\theta}_1$ & $\hat{\theta}_0$ & $\hat{\theta}_1$ \\
\hline
1 &    0 &    0 &    0 &    1 &    0 &    1 &    1 &    0 & 0.013 & 0.948 & 0.030 & 0.911 \\ 
  2 &    0 &    0 &    0 &    1 &    0 &    0 &    1 &    0 & 0.013 & 0.985 & 0.016 & 0.959 \\ 
  3 &    0 &    0 &    0 &    1 &    0 &    0 &    1 &    0 & 0.005 & 0.914 & 0.000 & 0.866 \\ 
  4 &    0 &    1 &    1 &    0 &    1 &    0 &    1 &    0 & 0.208 & 0.938 & 0.224 & 0.890 \\ 
  5 &    0 &    1 &    0 &    1 &    0 &    0 &    1 &    1 & 0.279 & 0.891 & 0.301 & 0.828 \\ 
  6 &    0 &    0 &    0 &    0 &    0 &    0 &    1 &    0 & 0.043 & 0.975 & 0.090 & 0.956 \\ 
  7 &    1 &    1 &    0 &    0 &    0 &    0 &    1 &    0 & 0.009 & 0.904 & 0.025 & 0.803 \\ 
  8 &    0 &    0 &    0 &    0 &    0 &    0 &    1 &    0 & 0.333 & 0.859 & 0.443 & 0.818 \\ 
  9 &    0 &    1 &    0 &    0 &    0 &    0 &    0 &    0 & 0.078 & 0.788 & 0.261 & 0.753 \\ 
  10 &    0 &    1 &    0 &    0 &    1 &    0 &    1 &    1 & 0.014 & 0.849 & 0.029 & 0.786 \\ 
  11 &    0 &    1 &    0 &    0 &    1 &    0 &    1 &    0 & 0.051 & 0.971 & 0.066 & 0.918 \\ 
  12 &    0 &    0 &    0 &    0 &    0 &    0 &    1 &    1 & 0.038 & 0.975 & 0.127 & 0.959 \\ 
  13 &    0 &    1 &    0 &    1 &    1 &    0 &    1 &    0 & 0.004 & 0.721 & 0.013 & 0.665 \\ 
  14 &    0 &    1 &    0 &    0 &    0 &    0 &    1 &    0 & 0.021 & 0.979 & 0.060 & 0.939 \\ 
  15 &    1 &    0 &    0 &    0 &    0 &    0 &    1 &    0 & 0.024 & 0.958 & 0.031 & 0.895 \\ 
  16 &    0 &    1 &    0 &    0 &    0 &    0 &    1 &    0 & 0.022 & 0.953 & 0.107 & 0.889 \\ 
  17 &    0 &    1 &    0 &    0 &    1 &    0 &    1 &    0 & 0.017 & 0.927 & 0.038 & 0.862 \\ 
  18 &    0 &    1 &    0 &    0 &    1 &    1 &    1 &    0 & 0.083 & 0.924 & 0.119 & 0.862 \\ 
  19 &    1 &    1 &    1 &    0 &    1 &    0 &    1 &    0 & 0.008 & 0.878 & 0.022 & 0.759 \\ 
  20 &    0 &    1 &    1 &    0 &    1 &    0 &    1 &    0 & 0.007 & 0.912 & 0.013 & 0.843 \\ 
\bottomrule
\end{tabular}
 \egroup
\vspace{2mm} 
\label{tbl:frac7}

{\it Mean, variance, and correlation estimation results for Fraction   data}
\begin{center}
\begin{center}
\begin{tabular}{c c c c c c c c c}
\toprule
& A1 & A2 & A3 & A4 & A5 & A6 & A7 & A8\\
\hline
 $\hat{ \mu}$ & -0.719 & 1.759 & 0.644 & 1.246 & -0.551 & 2.512 & 2.622 & 2.043 \\ 
$\Phi( \hat{ \mu})$  & 0.236 & 0.961 & 0.740 & 0.894 & 0.291 & 0.994 & 0.996 & 0.979 \\
 $\hat{ \sigma}^2$ & 22.687& 3.463 & 15.207 & 68.514 & 48.576 & 2.649 & 10.043 & 2.516 \\ 
\hline\\
\end{tabular}
\end{center}

\begin{tabular}{c c c c c c c c c}
\hline
$\hat{ \Sigma}^* $ & A1 & A2 & A3 & A4 & A5 & A6 & A7 & A8\\
A1 & 1.000 & 0.732 & 0.755 & 0.868 & 0.851 & 0.189 & 0.904 & 0.563 \\ 
  A2 & 0.732 & 1.000 & 0.712 & 0.656 & 0.756 & 0.318 & 0.809 & 0.680 \\ 
  A3 & 0.755 & 0.712 & 1.000 & 0.664 & 0.882 & 0.309 & 0.790 & 0.644 \\ 
  A4 & 0.868 & 0.656 & 0.664 & 1.000 & 0.787 & 0.174 & 0.883 & 0.345 \\ 
  A5 & 0.851 & 0.756 & 0.882 & 0.787 & 1.000 & 0.238 & 0.875 & 0.643 \\ 
  A6 & 0.189 & 0.318 & 0.309 & 0.174 & 0.238 & 1.000 & 0.272 & 0.271 \\ 
  A7 & 0.904 & 0.809 & 0.790 & 0.883 & 0.875 & 0.272 & 1.000 & 0.592 \\ 
  A8 & 0.563 & 0.680 & 0.644 & 0.345 & 0.643 & 0.271 & 0.592 & 1.000 \\ 
\bottomrule
\end{tabular}
\end{center}
	\label{frac-table}
\end{table}

\begin{figure}[h!]
 \centering\includegraphics[width=0.9\textwidth]{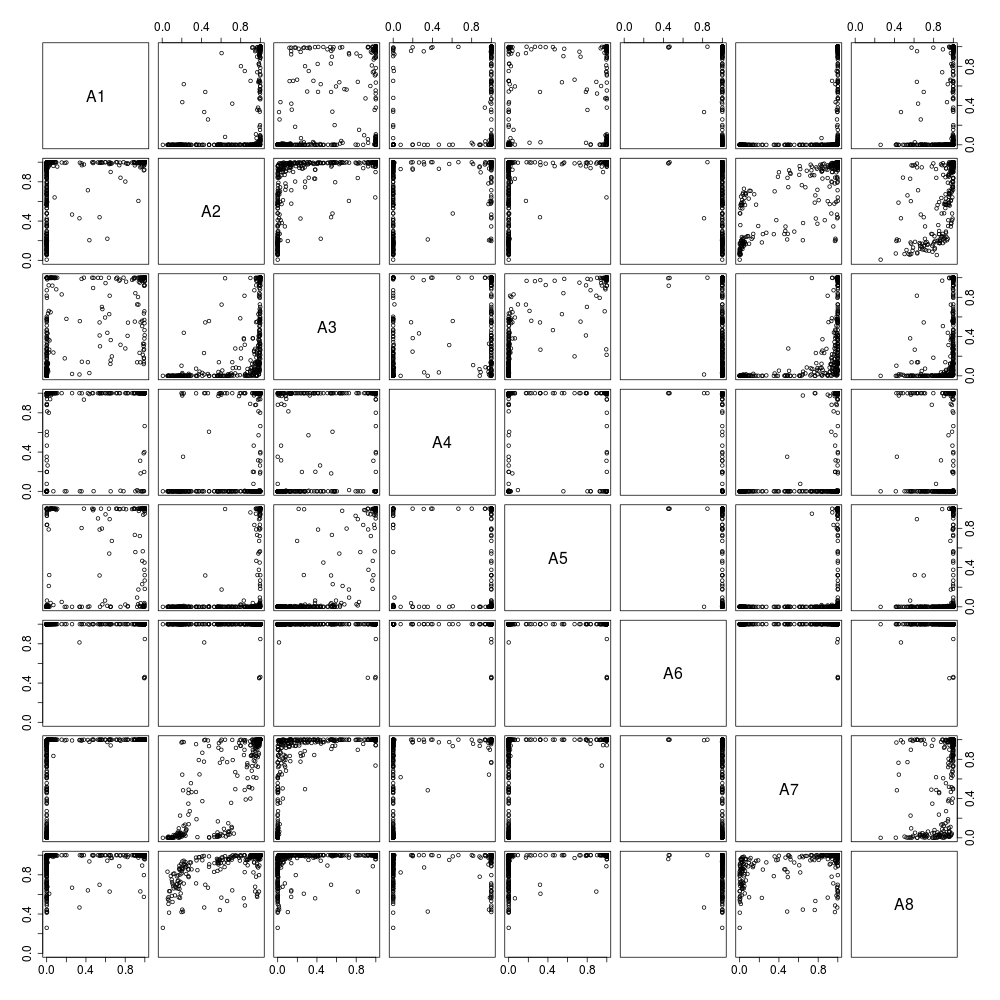}
 \caption{Scatter plots of      attribute scores for Fraction Subtraction data. }
 \label{frac-plot}
\end{figure}

 Figure \ref{frac-plot} shows   the plot of the estimated latent attribute mastery  scores from the PM-DINA model. We can see that some of the   plots (e.g, A1-A4, A1-A7 pairs) are similar to those in the simulation study when DINA model is true (upper panels in Figures \ref{fig3}), while others (e.g, A2-A3, A2-A8 pairs) seem to be more similar to that when the PM-DINA is true  (lower panels in Figures \ref{fig3}). 
In addition, most of  the subjects have the mastery scores of  A6   close to 1.
We also report in Table \ref{frac-table} the estimated mean, covariance, and correlation matrix from the PM-DINA model.  
  According to the discussions based on the simulation study,   larger values of the variances $\hat{\sigma}^2$'s serve as an indicator of an underlying DINA model while smaller $\hat{\sigma}^2$'s are likely the result from a PM-DINA model. 
 Interestingly, the results here show a mixture of both cases. For A1, A3, A4, A5 and A7, $\hat{\sigma}^2$'s  are large, while A2, A6 and A8 have relatively smaller $\hat{\sigma}^2$'s. 
   Moreover, we can see that most of the attributes are strongly-correlated, but A6 is weakly correlated with the others.   
Based on our results, we conclude   that the DINA model fits well on some attributes but not on others, while PM-DINA fits better overall. Hence, we  recommend  to use PM-DINA to fit the fraction subtraction  data.


\subsection{English Tests Data}
The second dataset is  the English tests dataset collected by the Examination for the Certificate of Proficiency in English (ECPE). Designed and organized by University of Michigan English Language Institute, the examination is developed to test high-level English language skills to determine   language proficiency of non-native speakers. It contains questions to evaluate grammar, vocabulary and reading skills. We consider     a subset of the data from 2003-2004 ECPE grammar section with 2922 subjects and 28 items  \citep{Liu}. The $Q$-matrix contains three attributes: Morphosyntactic Form, Cohesive Form and Lexical Form. 
 The data were also analyzed in \cite{chiu2016joint} using a joint likelihood estimation approach. 
 We again use DINA, GDINA and their partial mastery counterparts for fitting   this dataset. The information criteria results in Table 7 in the Supplementary Material \citep{supp} provide  strong evidence in favor of partial mastery \citep{kass1995bayes}. The PM-GDINA provides the best fit, yet the BIC difference between PM-DINA and PM-GDINA is negligible. 


%

\begin{table}[h!]
\caption{$Q$-matrix and estimation results. $\theta_{0,0}$ is for subjects with $\aalpha=(0,0,0)$, $\theta_{1,0}$ is for subjects only mastering the first of all required attributes, $\theta_{0,1}$ is for subjects only mastering the last of all required attributes, and $\theta_{1,1}$ is for   subjects mastering all required attributes;    note   here each item $j$    requires at most 2 attributes and thus has at most 4 different $\theta_{j,\aaa}$.} 
\centering
\bgroup
\def\arraystretch{1}
 \begin{tabular}{c c c c | c c c  c | c c c c}
\toprule
Item &\multicolumn{3}{c}{Q} & \multicolumn{4}{c}{PM-GDINA}  & \multicolumn{4}{c}{GDINA}\\
& Morph. & Coh. & Lex. & $\hat{\theta}_{00}$ & $\hat{\theta}_{10}$ & $\hat{\theta}_{01}$ & $\hat{\theta}_{11}$ & $\hat{\theta}_{00}$ & $\hat{\theta}_{10}$ & $\hat{\theta}_{01}$ & $\hat{\theta}_{11}$ \\
\hline
1 & 1 & 1 & 0 & 0.54 & 0.76 & 0.83 & 0.98 & 0.70 & 0.46 & 0.80 & 0.94 \\ 
  2 & 0 & 1 & 0 & 0.59 &  &  & 0.95 & 0.74 &  &  & 0.91 \\ 
  3 & 1 & 0 & 1 & 0.35 & 0.58 & 0.41 & 0.90 & 0.41 & 0.67 & 0.50 & 0.78 \\ 
  4 & 0 & 0 & 1 & 0.19 &  &  & 0.91 & 0.47 &  &  & 0.82 \\ 
  5 & 0 & 0 & 1 & 0.62 &  &  & 0.99 & 0.75 &  &  & 0.96 \\ 
  6 & 0 & 0 & 1 & 0.53 &  &  & 0.98 & 0.70 &  &  & 0.93 \\ 
  7 & 1 & 0 & 1 & 0.20 & 0.82 & 0.74 & 0.99 & 0.48 & 0.95 & 0.70 & 0.94 \\ 
  8 & 0 & 1 & 0 & 0.71 &  &  & 0.99 & 0.81 &  &  & 0.97 \\ 
  9 & 0 & 0 & 1 & 0.29 &  &  & 0.86 & 0.53 &  &  & 0.79 \\ 
  10 & 1 & 0 & 0 & 0.41 &  &  & 0.99 & 0.52 &  &  & 0.89 \\ 
  11 & 1 & 0 & 1 & 0.28 & 0.75 & 0.71 & 0.99 & 0.49 & 0.62 & 0.72 & 0.93 \\ 
  12 & 1 & 0 & 1 & 0.06 & 0.24 & 0.19 & 0.94 & 0.15 & 0.00 & 0.38 & 0.74 \\ 
  13 & 1 & 0 & 0 & 0.58 &  &  & 0.99 & 0.66 &  &  & 0.91 \\ 
  14 & 1 & 0 & 0 & 0.45 &  &  & 0.92 & 0.54 &  &  & 0.83 \\ 
  15 & 0 & 0 & 1 & 0.59 &  &  & 0.99 & 0.73 &  &  & 0.96 \\ 
  16 & 1 & 0 & 1 & 0.27 & 0.74 & 0.68 & 0.98 & 0.48 & 0.84 & 0.69 & 0.91 \\ 
  17 & 0 & 1 & 1 & 0.62 & 0.86 & 0.88 & 0.97 & 0.79 & 0.93 & 0.88 & 0.94 \\ 
  18 & 0 & 0 & 1 & 0.56 &  &  & 0.96 & 0.72 &  &  & 0.91 \\ 
  19 & 0 & 0 & 1 & 0.16 &  &  & 0.93 & 0.45 &  &  & 0.84 \\ 
  20 & 1 & 0 & 1 & 0.09 & 0.39 & 0.20 & 0.95 & 0.20 & 0.20 & 0.38 & 0.76 \\ 
  21 & 1 & 0 & 1 & 0.27 & 0.62 & 0.86 & 0.96 & 0.54 & 0.77 & 0.79 & 0.92 \\ 
  22 & 0 & 0 & 1 & 0.01 &  &  & 0.89 & 0.29 &  &  & 0.80 \\ 
  23 & 0 & 1 & 0 & 0.47 &  &  & 0.99 & 0.66 &  &  & 0.94 \\ 
  24 & 0 & 1 & 0 & 0.09 &  &  & 0.77 & 0.34 &  &  & 0.69 \\ 
  25 & 1 & 0 & 0 & 0.44 &  &  & 0.87 & 0.52 &  &  & 0.77 \\ 
  26 & 0 & 0 & 1 & 0.36 &  &  & 0.84 & 0.54 &  &  & 0.78 \\ 
  27 & 1 & 0 & 0 & 0.17 &  &  & 0.82 & 0.29 &  &  & 0.70 \\ 
  28 & 0 & 0 & 1 & 0.43 &  &  & 0.97 & 0.64 &  &  & 0.91 \\ 
\hline 
\end{tabular}

\begin{center}
{\it Covariance  and correlation estimation results for English   data}\vspace{0.02in}
\begin{tabular}{c c c c}
\hline
Cov.   & Morph. & Coh. & Lex.\\
Morph. & \textbf{1.126} & 0.867 & 1.010 \\ 
Coh. & 0.867 &  \textbf{1.083} & 0.973 \\ 
Lex. & 1.010 & 0.973 &  \textbf{1.337} \\ 
\hline
\end{tabular}
\hspace{1cm}
\begin{tabular}{c c c c}
\hline
Corr.  & Morph. & Coh. & Lex.\\
Morph. & 1.000 & 0.784 & 0.823 \\ 
Coh. & 0.784 & 1.000 & 0.809 \\ 
Lex.& 0.823 & 0.809 & 1.000 \\ 
\hline
\end{tabular}
\end{center}
  \vspace{2mm}
  \label{tbl:eng}
 \egroup
\end{table}

\begin{figure}[h!]
 \centering\includegraphics[width=0.9\textwidth]{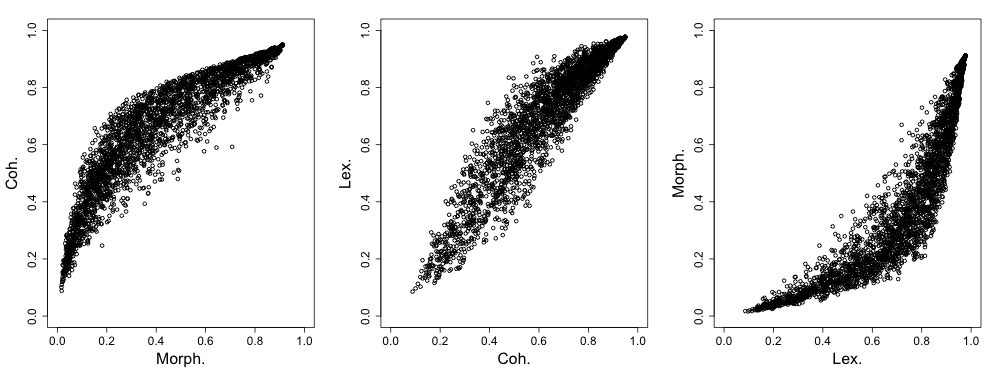}
 \caption{Estimated attribute scores    by PM-GDINA for   English   data. }
 \label{english-fig}
\end{figure}

Table \ref{tbl:eng}  provides the estimated $\theta_{j,\aaa}$'s from the GDINA and the PM-GDINA. We observe that the estimates from the PM-GDINA satisfy the monotonicity assumption while the results for items 1 and 12 from the GDINA violate that. In addition,   the PM-GDINA generally gives smaller estimates of $\hat\theta_{j,\mathbf 0}$ and larger estimates of $\hat\theta_{j,\mathbf 1}$  than the GDINA, which agrees with the simulation study when the PM-GDINA is correct.
 For the population parameters, the estimated $ \mu$'s from the PM-GDINA is $(0.227,  0.624, 0.944)$ which corresponds to $(0.590, 0.734, 0.827)$ on the 0-1 scales of the mastery scores. Comparing to the fraction subtraction data, all estimated $\mu$'s are relatively small; moreover, all diagonal terms of estimated covariance matrix in Table \ref{tbl:eng} are fairly close to one. These results  suggest that the PM-GDINA may be a better fit for all attributes than the GDINA. 
 The plots of the attribute mastery scores in Figure \ref{english-fig} further  demonstrate this.   Instead of concentrating at the edges and corners, the  majority   of the subjects' mastery scores lie    in the middle   of the plot. According to the simulation results  (e.g., Figure \ref{fig3}), this suggests that PM-GDINA fits the data better than the GDINA model. Overall, the results provide strong evidence in favor of partial mastery over binary mastery for all items and attributes. 
 
\begin{figure}[h!]
\centering
\begin{minipage}{1\textwidth}
 \centering
\includegraphics[width=0.45\textwidth]{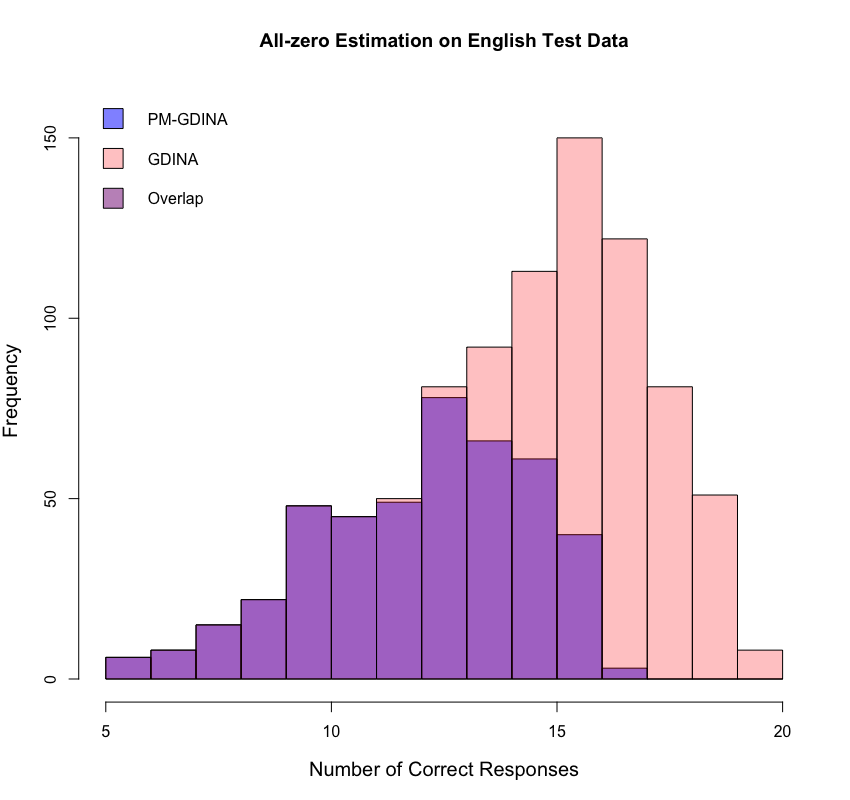}
\includegraphics[width=0.45\textwidth]{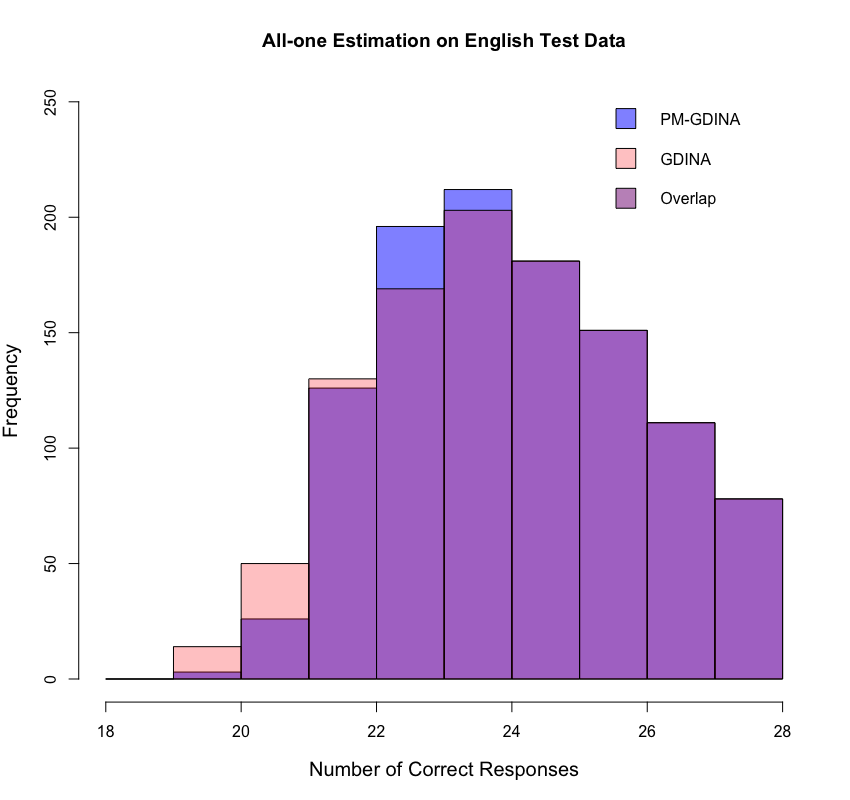}
 \end{minipage}
 \caption{ (Left) Frequency of estimated all-zero $\aalpha$ of English Test Data from PM-GDINA and GDINA, versus the number of correct responses. (Right) Frequency of estimated all-one $\aalpha$ from PM-GDINA and GDINA, versus the number of correct responses.}
 \label{fig-Eng}
\end{figure}

  We further investigate goodness-of-fit of the GDINA and PM-GDINA on individual levels. 
  In Figure \ref{fig-Eng}, we present the frequency of estimated all-zero $\aaa$ (left panel) or all-one $\aaa$ (right panel) attribute profiles -- students who have not mastered any skill or have mastered all three skills, respectively -- from PM-GDINA and GDINA versus the number of correct responses. 
The right panel suggests that the two models have similar classification performance for the classification of students into the all-one attribute profile. However, the left panel indicates a substantial difference for the classification of students into the zero skill mastery profile. Thus, while the PM-GDINA and GDINA models are similar in classifying students with less than 12 correct responses into the all-zero profile, GDINA also estimates zero skill mastery for a substantial proportion of students who had more than 12 correct responses,  while PM-GDINA does not. 
  Take subject 1792 as an example, who got correct answers in 20 out of the total 28 questions (Table \ref{tbl:eng7}), 
  the GDINA classifies its attribute pattern as $(0,0,0)$ with its posterior probability  (0.48)     significantly higher  than those of other attribute profiles, which indicates the test-taker does not master any of the attributes. On the other hand, the PM-GDINA estimates   $\hat{\mathbf d} = (0.41,  0.65, 0.67)$ suggest partial mastery of all the three skills. 
 In   Table  \ref{tbl:eng7}, we also list observed responses from the other four subjects who answered at least 18 out of 28 items correctly but were classified as $(0,0,0)$ by the G-DINA model.
  On the other hand, the PM-GDINA estimates reflect partial-mastery of the three skills. 
  We also present in Table \ref{tbl:eng7} the posterior estimation result of each attribute from the GDINA (values in   parentheses of the column ``GDINA Class").  Though subject  1792's results are more similar to the PM-GDINA, the rest four subjects' results are quite different. The posterior estimate of each attribute of these four subjects from GDINA appears close to zero, while each of subjects answered 18 out of 28 items correctly.  
  
 Overall,   the  above analysis   shows that the PM-GDINA appears to statistically fit the data better.  Practically, we may further consult education domain experts to decide which results should be used in reporting the cognitive diagnosis results for the subjects.

  One interesting observation pointed out by an anonymous  reviewer is that the subject 1792 correctly answered 10 out of those 13 items that require the first attribute, seemingly contradicting the estimation result of a relatively low mastery score $d_1= 0.41$. 
  One reason for this is that  the estimation of each attribute is largely affected by the item parameters. Consider items 13 and 27 for example. Though both items only require the first attribute, we can see from the estimation results in Table \ref{tbl:eng} that item 13 is a much easier problem than item 27; for a subject without mastering any attribute, $\dd =(0,0,0)$, he/she still has a probability of $\theta_{000} = 0.58$ to correctly answer item 13, but only a probability of $\theta_{000} = 0.17$ to correctly answer item 27. Such differences in item parameters would lead to different estimated mastery scores even for subjects answering the same number of items correctly;  for instance,  conditional on the other items' responses, answering item 13 correctly but not item 27 (as did by subject 1792) would result in a lower $d_1$ estimate than  the other way around. 
 Another major  reason  is that  the estimation of each attribute also depends on the $\qqq$-vectors. Note that
 several items requiring the first attributes also require the other attributes; under the compensatory GDINA and PM-GDINA model assumption,   the absence of one required attribute  can be compensated by the presence of other latent attributes. Take item 1, which requires the first two attributes,  for   example; from the estimation results, an individual with attribute profile (010), though lacking the first attribute, still has a probability of 0.83 to answer this item correctly.  Thus, the simple subscore of counting how many items requiring each attribute were correctly answered often does not provide a reasonable estimation of the latent attribute, a phenomenon  commonly occurring  in multidimensional latent variable models, such as  CDMs \citep{Rupp} and item response theory models \citep{reckase2009multidimensional}. This is also the reason why latent variable statistical models are often preferred in educational measurement rather than simply using  total scores or subscores. 
 
\begin{figure}[h!]
\centering
\begin{minipage}{1\textwidth}
 \centering
 \includegraphics[width=0.45\textwidth]{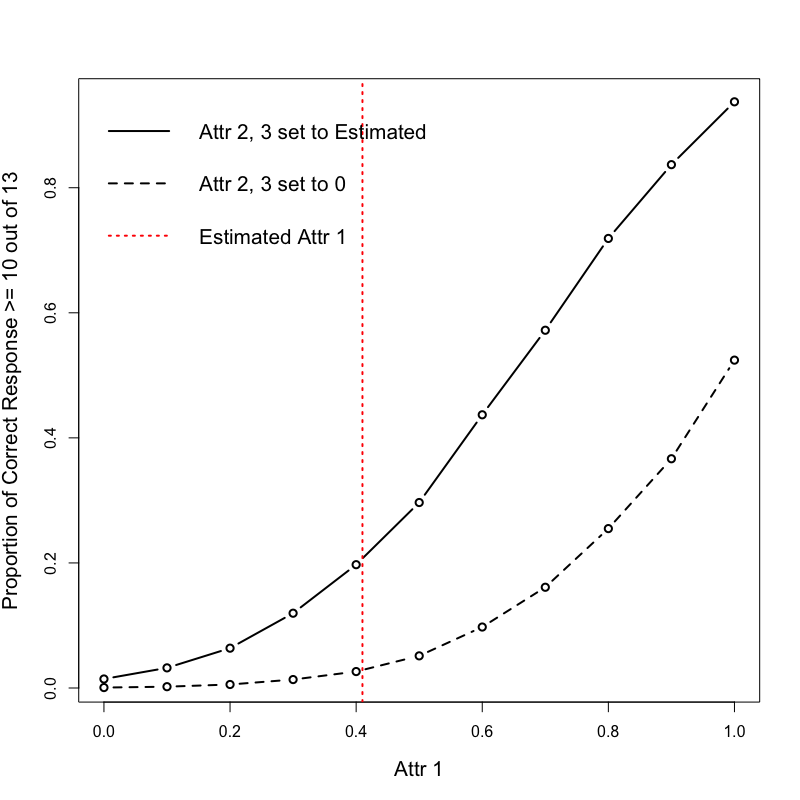}
\end{minipage}
 \caption{The estimated probability (y-axis) of providing correct answers to at least 10 out of those 13 items requiring attribute 1, for mastery scores $\hat{\mathbf d} = (d_1,  0.65, 0.67)$ (solid curve) and $\hat{\mathbf d} = (d_1,  0, 0)$ (dashed curve), with first attribute's mastery score $d_1$ (x-axis) varying from 0 to 1. The dotted vertical line indicates the estimated  score $d_1=0.41$ of subject~1792. }
 \label{fig-1792}
\end{figure}

 We use a simulation study  to further illustrate this. Under the PM-GDINA model, using the data $Q$-matrix and the estimated item parameters in Table \ref{tbl:eng}, we  simulate 10,000 independent response vectors to the 28 items from the estimated mastery score $\hat{\mathbf d} = (0.41,  0.65, 0.67)$ of subject 1792. We find that about 21\% out of  the 10,000 simulations have correct  answers to at least 10 out of those 13 items that require the first attribute, indicating that the seemly contradictory observation is a normal phenomenon under the estimated model parameters. 
We further present in Figure \ref{fig-1792} the estimated probability of providing correct answers to at least 10 out of those 13 items for mastery scores $\hat{\mathbf d} = (d_1,  0.65, 0.67)$ with the mastery scores of attributes 2 and 3 taking the same estimated values as subject 1792 (solid curve) and $\hat{\mathbf d} = (d_1,  0, 0)$ (dashed curve), where the mastery score of the first attribute $d_1$ varies from 0 to 1. 
Comparing the two curves, we can see that when attributes 2 and 3 take 0 values, the chance of observing at least 10 correct answers out of those 13 items is much lower. In addition, both curves increase with $d_1$, indicating a  positive correlation between the number of correct answers and the level of the first mastery score.
All these results suggest  the reasonableness  of the estimation result for subject 1792.

 \begin{table}[h!]
 \caption{Individual case study for English  Data. ``GDINA Class" column gives the estimated attribute profiles by GDINA and the posterior estimate of each attribute (in  parentheses), and ``PM-GDINA" column gives the estimated mastery scores. }
\begin{center}
	
 \begin{tabular}{c c c c  }
\hline
Subject & Response & GDINA Class & {PM-GDINA Scores} \\
\hline
1792 & 1100111011111111110010111100  & 000 (0.36, 0.45, 0.51) 
&   (0.41, 0.65, 0.67) \\ 
\hline
  429&  1010111111001010010111111001 & 000 (0.05, 0.20, 0.32)  
 &   (0.28, 0.56, 0.63)
\\ 
  824& 1110011101011001101111011101 & 000 (0.23, 0.20, 0.34)    &   (0.41, 0.62, 0.65)
\\ 
  1366& 1001010011111101011010111011 & 000 (0.02, 0.02, 0.05)    &   (0.36, 0.51, 0.51) \\ 
2273  &1111110111001110110110011011 & 000 (0.08, 0.17, 0.19) 
 &   (0.38, 0.62, 0.65) 
 \\ 
\hline
\end{tabular}
  \end{center}
    \label{tbl:eng7}
  \end{table}

  Finally, we would like to point out that the hierarchical structure of the latent attributes for this dataset was  studied in \cite{templin2014hierarchical} under CDMs.
 Even though the hierarchy structure is not used in the CDM and PM-CDM modeling of this work, the corresponding hierarchical models can be viewed as submodels of the fitted CDM and PM-CDM by constraining the model parameters to respect the hierarchy.
As the sample size of this dataset is relatively large ($N=2922$), fitting the general models would yield similar results as the hierarchical submodels.
This was also found in the data analysis in \cite{templin2014hierarchical}  that
``{\it Agreement between the two models [the hierarchy and general models] was very high, providing matching classifications for 93.6\% of examinees. The two models provide similar classifications of examinees with only
a few cases where classifications differed substantially.}"
For this reason, this work focuses on the general CDM and PM-CDM.

\section{Discussion}
In this paper, we propose a mixed membership modeling approach for cognitive diagnosis, PM-CDM, which allows for partial mastery along multiple dimensions of latent attributes. The proposed approach covers CDMs as special cases and provides a more flexible tool for cognitive diagnosis assessment. 
We develop a Bayesian estimation method that is applicable to all PM-CDMs and demonstrate accurate parameter recovery in simulated setting with partial mastery assumption. Using simulation studies, we investigate the impact of model misspecification when the true data generation process assumes binary mastery but partial mastery models are fitted to the data, and vice versa. Our results indicate that partial mastery models are able to fit simulated data better than binary mastery CDMs when partial mastery is present, and provide comparable (or slightly worse) results to CDMs when binary mastery is the correct assumption. In addition to goodness-of-fit measures such as AIC and BIC, we suggest several diagnostics that could be used for determining when partial mastery models are more appropriate. In the two real data examples considered in this paper, we find strong evidence in support of partial mastery. For the fraction subtraction data, partial mastery was clearly more appropriate than binary mastery for some attributes (such as A3 ``simplify before subtracting'') but not for others (such as A7 ``subtract numerators''), whereas in the English tests data, partial mastery specification was found to be more appropriate for all attributes. Examining discrepancies between estimated individual-level partial mastery scores and attribute profiles, we observe that PM-CDMs provide more detailed information about the skills learned that skills that need study. This property of PM-CDMs has the potential to lead to more effective remediation programs.

The proposed Gaussian copula modeling approach   needs $K+K(K+1)/2$ parameters to model the distribution of  the attribute mastery score $\dd$. 
The computation cost of the proposed Gibbs sampling method increases quadratically with respect to $K$.    
 An interesting   direction is to   further reduce the model complexity, especially when the number of attributes $K$ is large.  We may consider a factor analysis model for the latent scores; for instance, the one-factor model is
$
\{\Phi^{-1}(d_k); k=1,\cdots,K\}^\top= \mu + \Lambda \eta +\epsilon,
$
where $\Lambda$ is a $K\times 1$ loading vector, $\eta$ is a subject specific factor following standard normal distribution and $\epsilon$ follows a multivariate $K$-dimensional normal distribution with zero mean and diagonal covariance $E$. Then this model  corresponds to 
 $\{\Phi^{-1}(d_k); k=1,\cdots,K\}^\top\sim N( \mu,\Lambda\Lambda^\top  +E),$ 
which has fewer model parameters than the proposed  model in Section 2 (when $K\geq 4$). 
 The above factor analysis modeling is similar  to the higher-order factor models and higher-order CDMs   proposed in \cite{dela}, \cite{templin2008robustness},  and \cite{culpepper2019development}; this is also   related  to the multidimensional IRT model \citep{reckase2009multidimensional}. 
 However, as discussed in Section 3.1, our method differs from these existing studies due to the different cognitive diagnosis modeling from the latent scores to responses. 
Alternatively,  to reduce the computation cost when $K$ is large, we may use the variational approximation inference method \citep{blei2003latent,Erosheva2007,blei2007correlated} or moment based method \citep{zhao2017fast} for PM-CDMs.

 Another interesting extension of this study is to allow for mixtures of binary and partial mastery across    attributes. As illustrated in the analysis of the Fraction Subtraction data, in practice some attributes may satisfy the partial mastery assumption while other may follow the binary assumption. 
 Moreover, as discussed in the analysis of the English data, the extension to  hierarchical PM-CDMs is also worthy of further investigation. Such an extension would need to carefully incorporate prespecified hierarchical structures among the latent attributes by domain experts. 
 
 Finally, in this paper, we have assumed that the pre-specified $Q$-matrix is known and correct. In practice, the $Q$-matrix is usually constructed by the users and may not be accurate.  
A misspecified $Q$-matrix could lead to substantial lack of fit and, consequently, erroneous classification of subjects \citep{Rupp20082,dela2}.
 Various methods for $Q$-matrix validation and estimation have been proposed under CDMs   \citep*[e.g.,][]{JLGXZY2012,JLGXZY2011,decarlo2012recognizing,Chen2014,de2015general,Xu2017,chen2018bayesian}.
 It would be interesting to use these approaches to develop a more general PM-CDM framework for estimating the $Q$-matrix of multiple latent dimensions.

   \section*{Acknowledgements}
  The authors thank the editors, an associate editor, and two anonymous referees for their constructive comments.  
 This research is partially supported by NSF grants  CAREER SES-1846747,  SES-1659328 and DMS-1712717, and IES
grant R305D160010. Elena Erosheva would like to acknowledge that this research has partly taken place while she was a visiting professor at the Laboratorie J. A. Dieudonné, Université Côte d‘Azur, CNRS, Nice, France.

 \begin{supplement}[id=suppA]
 \slink[doi]{COMPLETED BY THE TYPESETTER}
 \sdatatype{.pdf}
  \stitle{Supplement to ``Partial-Mastery Cognitive Diagnosis Models"}
 \sdescription{The supplementary material includes details for the   Gibbs sampling algorithm and additional simulation    and   data analysis results.}
 \end{supplement}
%

\bibliographystyle{chicago}
\bibliography{bibEduc.bib}
 
\end{document}